\documentclass[%
 aps,
 ph,%
 amsmath,amssymb,
 reprint,%
t]{revtex4-1}
\usepackage[utf8]{inputenc}
\usepackage[T1]{fontenc}
\usepackage[colorlinks,citecolor=blue,urlcolor=blue,linkcolor=blue]{hyperref}
\usepackage{graphicx}
\usepackage{dcolumn}
\usepackage{bm}
\usepackage{amsmath}
\usepackage{amsfonts}

\usepackage{titlesec}
\titlespacing*{\section}{0pt}{3.5ex}{3.5ex}
\titlespacing*{\subsection}{0pt}{3.5ex}{3.5ex}
\titlespacing*{\subsection}{0pt}{3.5ex}{3.5ex}
\titlespacing*{\subsubsection}{0pt}{3.5ex}{3.5ex}

\usepackage{amssymb}
\usepackage{mathrsfs}
\usepackage{braket}
\usepackage{bbm}
\usepackage{xcolor}
\definecolor{olive}{rgb}{0.3, 0.4, .1}
\definecolor{fore}{RGB}{249,242,215}
\definecolor{back}{RGB}{51,51,51}
\definecolor{title}{RGB}{255,0,90}
\definecolor{blackViolet}{RGB}{138,43,226}
\definecolor{dgreen}{rgb}{0.,0.6,0.}
\definecolor{gold}{rgb}{1.,0.84,0.}
\definecolor{JungleGreen}{cmyk}{0.99,0,0.52,0}
\definecolor{blackGreen}{cmyk}{0.85,0,0.33,0}
\definecolor{RawSienna}{cmyk}{0,0.72,1,0.45}
\definecolor{Magenta}{cmyk}{0,1,0,0}
\definecolor{wood}{RGB}{139,115,85}
\definecolor{dorange}{RGB}{255,127,0}
\definecolor{dolive}{RGB}{85,107,47}
\definecolor{drg}{RGB}{255,165,0}

\DeclareMathAlphabet{\mathpzc}{OT1}{pzc}{m}{it}

\usepackage{multirow}

\usepackage{tikz}
\usepackage{colortbl}
\usepackage{multimedia}
\usepackage{xcolor}

\usepackage{graphicx}
\usepackage{ragged2e}
\usepackage{array}
\usepackage{rotating}
\usepackage{longtable}
\usepackage{arydshln}
\usepackage{multirow}

\newtheorem{lemma}{Lemma}[section]

\usepackage{simplewick}
\usepackage{dirtytalk}

\usepackage{mathtools}

\usepackage{stmaryrd}
\usepackage{geometry}
\begin{document}


\title[]{Natural-orbital representation of molecular electronic transitions}%

\author{Thibaud Etienne}
\email{thibaud.etienne@univ-lorraine.fr}%
\affiliation{%
Université de Lorraine, CNRS, LPCT, F-54000 Nancy, France%
}%

\date{\today}%

\begin{abstract}
\noindent This paper aims at introducing the formal foundations of the application of reduced density-matrix theory and Green's function theory to the analysis of molecular electronic transitions. For this sake, their mechanics, applied to specific objects containing information related to the passage and the interference between electronic states — the difference and the transition density operators — are rigorously introduced in a self-contained way. After reducing the corresponding $N$-body operators (where $N$ is the number of electrons in the system) using an operator partial-trace procedure, we derive the kernel of the reduced one-body difference and transition density operators, as well as the matrix representation of these operators in a finite-dimensional one-particle-state basis. These derivations are done in first and second quantization for the sake of completeness — the two formulations are equivalently present in the literature — and because second quantization is extensively used in a second part of the paper. Natural orbitals are introduced as appropriate bases for reducing the dimensionality of the problem and the complexity of the analysis of the transition phenomenon. Natural-orbital representation of density operators are often used as a tool to characterize the nature of molecular electronic transitions, so we suggest with this contribution to revisit their theoretical foundations in order to better understand the origin and nature of these tools. \\ $\;$ \\  \textit{Keywords: Molecular excited states electronic-structure theory; Reduced density matrix theory; Natural-orbital representation of density operators.}

\end{abstract}

\maketitle

\section{Introduction}\label{sec:intro}
\noindent Light-matter interaction is ubiquitous in nature. Light capture or emission phenomena have been extensively discussed in the scientific literature for years. In particular, excited-state electronic-structure theory has taken an increasingly advantageous place in the theoretical community. However, the process by which an electromagnetic wave interacts with the electronic cloud of a physical system is very complex, and plenty of methods are proposed for computing the excited states of molecules. For molecular systems, they can be derived using the wavefunction \cite{maurice_configuration_1995-1,david_sherrill_configuration_1999,sekino_linear_1984,koch_coupled_1990} , the electron density \cite{hirata_configuration_1999,casida_time-dependent_1995,ziegler_derivation_2014,fromager_individual_2020}, the one-body reduced density matrix \cite{pernal_time-dependent_2007-1}, or the Green function \cite{rebolini_electronic_2013,leng_gw_2018,oddershede_polarization_1978,strinati_application_1988} as a central object of interest. Moreover, the light-matter interaction \textit{model} itself — how does light interact with matter — is the object of many fascinating scientific discussions.

Amongst the quantities and objects of interest in this framework, one obviously thinks about the electronic transition energy. Indeed, it is a difficult task to accurately reproduce and predict electronic spectral properties: optical absorption/emission frequency and relative intensity, band shapes, etc. This becomes tremendously complicated when effects of the chromophores surroundings (solvent, interface, amino acids, nucleic acids, etc.) and chromophore-environment dynamics come into play. What will retain our attention in this contribution is the way light-matter interaction polarizes the electronic cloud of a molecular system: how is the electronic structure of a molecule reorganized upon light absorption or emission?

The model for such a description varies according to the nature of the system of interest — what happens in solid state and in finite-dimensional systems is not yet described in a fully unified way, though tentatives for reconciliating the two worlds start to arise \cite{bappler_exciton_2014} — and according to the framework for the excited-state computation.

In this context, a bench of tools is available for performing qualitative \cite{dreuw_single-reference_2005,luzanov_charge_1978,plasser_detailed_2017,luzanov_structure_1980,luzanov_analysis_2006,ronca_charge-displacement_2014,li_particlehole_2015,etienne_charge_2019-1} and quantitative analyses \cite{monino_upper_2021,etienne_toward_2014,plasser_new_2014-2,breuil_diagnosis_2019-2,luzanov_interpretation_1980,ciofini_through-space_2012,etienne_charge_2019-1} of molecular electronic transitions. The most used ones are one-electron density functions and one-electron wavefunctions (spinorbitals), for they are easy to manipulate theoretically, and to visualize. In this contribution, we will revisit the basic foundations of two increasingly used analysis paths: one takes its origin in the one-electron difference density matrix \cite{head-gordon_analysis_1995,plasser_new_2014-2,dreuw_single-reference_2005,etienne_comprehensive_2020-1}, and the other one arises from the definition of a one-electron transition density matrix \cite{luzanov_application_1976,dreuw_single-reference_2005,plasser_new_2014-2,etienne_comprehensive_2020-1}. Since we think there exists vocabulary ambiguities relatively to these two frameworks — what is the difference between the one-electron reduced (difference or transition) density operator, kernel, matrix, and function? — we will reconstruct all these objects from the very fundamental density-operator and Green's function scheme.

A discussion will follow relatively to the physical content of these objects, and a comparison will be established between the two characterization paths — what does a tool provide that cannot be derived from the other one? What are the interpretation limits of these objects? In particular, we will focus much of our attention on the role of natural transition orbitals \cite{martin_natural_2003} in molecular quantum mechanics.

With this, we hope to provide the reader with a clear and exhaustive picture of the difference and transition reduced-quantity frameworks. Technical details are intentionally brought to the reader for the sake of completeness and for facilitating self-derivation.

\newpage
\section{Hypotheses and notations} \label{sec:hyp}
\noindent In this contribution we will consider $N$--electron molecular systems. For any electronic quantum state considered below, the $N$ electrons are distributed in $K$ complex-valued one-electron wavefunctions ($K \geq N$) defined on $S_4 \coloneqq \mathbb{R}^3\times\left\lbrace \uparrow , \downarrow \right\rbrace $, where the ``up'' ($\uparrow$) and ``down'' ($\downarrow$) arrows denote $\alpha$ and $\beta$ electron spin projections. 

We therefore define two integer intervals:
\begin{align*}
I &\coloneqq \llbracket 1,N\rrbracket,\\
C &\coloneqq \llbracket 1,K\rrbracket.
\end{align*}
Let the $N$-electron system be decomposed into two subsystems: a one-electron subsystem $S_i$ (with $i\in I$), and a $(N-1)$-electron subsystem $S_{I\textbackslash \left\lbrace i\right\rbrace}$. As we will see, the choice of $i$ is purely arbitrary due to the fact that electrons are indistinguishable. Let $\mathcal{S}_i$ be a $K$-dimensional state space corresponding to $S_i$ with 
$$\textbf{K} \coloneqq (\ket{k_p})_{p\in \llbracket 1,K\rrbracket}$$ 
a one-particle-state orthonormal basis of $\mathcal{S}_i$. Its linear span, $\mathcal{K}$, is defined as
\begin{equation*}
\mathcal{K} \coloneqq \mathrm{span}\left(\textbf{K}\right) = \left\lbrace \sum _{q = 1}^{K} \lambda_q\, \ket{k _q} \; : \; \left(\lambda _q\right)_{q\in C} \in \mathbb{C}^{1\times K}\right\rbrace.
\end{equation*}
 Let $\mathcal{S}_{I\textbackslash \left\lbrace i\right\rbrace}$ be a $W$-dimensional state space corresponding to $S_{I\textbackslash \left\lbrace i\right\rbrace}$, with $$\textbf{W} \coloneqq (\ket{w_p})_{p\in \llbracket 1,W\rrbracket}$$ an $(N-1)$-particle-state orthonormal basis of $\mathcal{S}_{I\textbackslash \left\lbrace i\right\rbrace}$, and $\mathcal{W}$ the linear span of $\textbf{W}$. We have that
 $$\forall i\in I,\, \mathcal{K} = \mathcal{S}_i \; \mathrm{and}\; \mathcal{W} = \mathcal{S}_{I\textbackslash \left\lbrace i\right\rbrace}.$$
In this contribution we will consider an orthonormal basis of $(M+1)$ molecular electronic states,
$$\textbf{S} \coloneqq \left(\ket{\psi _m}\right)_{m\in \llbracket 0, M\rrbracket}.$$ 
We then set the integer interval corresponding to the numbering of the states in the $\textbf{S}$ basis:
\begin{equation*}
S \coloneqq \llbracket 0, M\rrbracket
\end{equation*}
and we define $\mathcal{S}$ as the linear span of $\textbf{S}$. We also assume that the $N$-electron quantum states in \textbf{S} satisfy
$$\forall i \in I, \,\forall q \in S, \, \ket{\psi _q} \in \mathcal{S}_i \otimes \mathcal{S}_{I\textbackslash \left\lbrace i\right\rbrace}$$
with $\mathcal{S}_i \otimes \mathcal{S}_{I\textbackslash \left\lbrace i\right\rbrace} = \mathrm{span}\left(\ket{k_p}\otimes \ket{w_q}\right)_{(p,q)\in \llbracket1,K \rrbracket \times \llbracket 1, W\rrbracket}.$
\section{One-body reduced difference and transition density objects}\label{sec:AppRDMs}
\noindent For every $(\ell,m)$ in $S^2$, we set the $N$-body state difference
$$\hat{\Delta}_N^{\ell\rightarrow m} \coloneqq \ket{\psi _m}\bra{\psi _m} - \ket{\psi _\ell}\bra{\psi _\ell}$$
and the $N$-body transition (sometimes called \textit{state-transfer})
$$\hat{\mathrm{T}}_N^{\ell\rightarrow m} \coloneqq \ket{\psi _m}\bra{\psi _\ell}$$
density operators. 
Both the $N$-body state difference and the $N$-body transition density operators are linear maps from the space of $N$-electron states to itself. Note that the latter could be rewritten $\hat{\mathrm{T}}_N^{\ket{\psi_\ell}\rightarrow \ket{\psi_m}}$ for preventing any confusion with its dual map acting in $\mathcal{S}^*$, i.e.,
$$\ket{\psi _\ell}\bra{\psi _m} = \hat{\mathrm{T}}_N^{\bra{\psi_\ell}\rightarrow \bra{\psi_m}} = \left(\hat{\mathrm{T}}_N^{\ket{\psi_\ell}\rightarrow \ket{\psi_m}}\right)^\dag$$
provided that there is a bra associated with each ket in $\mathcal{S}$. The choice to work in $\mathcal{S}$ or in $\mathcal{S}^*$ is arbitrary and author-dependent — though the choice is usually not made explicit. Note however that $\mathcal{S}$ is a space of quantum-state vectors, while $\mathcal{S}^*$ is a space of linear functionals, which justifies our initial convention. For the sake of readibility, we will keep our original notation.
 
 The difference density operator has a matrix representation in the $\textbf{S}$ basis,
$$\bm{\Delta}_N^{\ell\rightarrow m} \coloneqq \mathcal{M}\left( \hat{\Delta}_N^{\ell\rightarrow m} , \textbf{S}\right),$$
that is diagonal, with
\begin{align*}
\forall (r,s)\in S^2, \,  (\bm{\Delta}_N^{\ell\rightarrow m})_{r,s} &=\braket{\psi _r | \hat{\Delta}_N^{\ell\rightarrow m} | \psi _s} \\&= \delta_{r,s}\left(\delta_{s,m} - \delta_{s,\ell}\right),
\end{align*}
where $\delta_{\cdot , \cdot}$ is Kronecker's delta. In other words, the only two non-zero $N$-\textit{body-states transition occupation numbers} are $(+1)$ for the arrival state, and $(-1)$ for the departure state. \par In what follows, we will expose the machinery to reach, in first and second quantization, the one-body reduced difference and transition density operators, kernel, and matrix representations in the \textbf{K} basis. Since the machinery in first quantization is common for the two types (difference and transition) of operators, we will provide a single derivation procedure, and the fact that it can be applied to both difference and transition operators will be highlighted by the $\lambda$ symbol: if one replaces $\lambda$ by $\Delta$ (respectively, $\mathrm{T}$) in what follows, the procedure leads to the one-body reduced difference (respectively, transition) density operators, kernel and matrix representation in the \textbf{K} basis.
\subsection{First quantization}
\noindent Given a special — arbitrary — numbering of the electrons, the partial trace of the $N$-body difference ($\lambda = \Delta$) and transition ($\lambda = \mathrm{T}$) density operators where contributions of $(N-1)$ electrons are traced out, leaving only the contribution of the $i^\mathrm{th}$ electron, reads
$$\hat{\lambda}_{1,i}^{\ell\rightarrow m} \coloneqq \mathrm{tr}_{\mathcal{S}_{I\textbackslash \left\lbrace i\right\rbrace}}\left( \hat{\lambda}_N^{\ell\rightarrow m}\right).$$
Since the $N$ electrons are undistinguishable, there exist $N$ such decompositions. The \textit{total} one-electron difference and transition density operators are both a sum of every single-electron contribution
$$
\hat{\lambda}_1^{\ell\rightarrow m} \coloneqq \sum _{i \in I}\mathrm{tr}_{\mathcal{S}_{I\textbackslash \left\lbrace i\right\rbrace}}\left(\hat{\lambda}_N^{\ell\rightarrow m}\right). $$
In order to understand the meaning of the partial trace in the context of density operators analysis, we apply it to any $N$-body pure state difference and transition density operator. Again, the $N$-body system can be decomposed into two electron subsystems $S_i$ and $S_{I\textbackslash\left\lbrace i\right\rbrace}$ (with $i\in I$). The $N$-electron spatial-spin coordinates
$$(\textbf{s}) \coloneqq (\textbf{s}_i)_{i\in I}$$
can be separated into a one-electron tuple of spatial-spin coordinates for the $i^\mathrm{th}$ electron, $(\textbf{s}_i)$, and an $(N-1)$-electron tuple of spatial-spin coordinates $\left(\textbf{s}_{-i}\right)$, where
$$\left(\textbf{s}_{-i}\right) \coloneqq (\textbf{s}_j)_{j \in I \textbackslash\left\lbrace i\right\rbrace}.$$
Consider $(1 < i < N)$. Then,
$$(\textbf{s}_{j})_{j \in I \textbackslash\left\lbrace i\right\rbrace} = \left(\textbf{s}_1, \ldots, \textbf{s}_{i-1} , \textbf{s}_{i+1}, \ldots, \textbf{s}_N\right),$$
and the product of any $N$-electron wavefunction value with the complex conjugate of any other $N$-electron wavefunction value, e.g., for any $(a,b)$ in $S^2$,
$$\psi _a(\textbf{s}_1,\ldots , \textbf{s}_i, \ldots, \textbf{s}_N) \psi _b^*(\textbf{s}_1',\ldots,\textbf{s}_i',\ldots, \textbf{s}_N'),$$
is equal to 
\begin{align*}
&(-1)^{i-1}\psi _a(\textbf{s}_i, \textbf{s}_1,\ldots ,\textbf{s}_{i-1} , \textbf{s}_{i+1} \ldots, \textbf{s}_N) \\ &\times (-1)^{i-1}\psi _b^*(\textbf{s}_i',\textbf{s}_1',\ldots,\textbf{s}_{i-1}', \textbf{s}_{i+1}',\ldots, \textbf{s}_N'),
\end{align*}
i.e., since $(-1)^{i-1}(-1)^{i-1} = (-1)^{2(i-1)} = 1$,
$$\psi _a (\textbf{s}_i, \textbf{s}_{-i})\psi_b^* (\textbf{s}_i',\textbf{s}_{-i}').$$
This naturally extends to ($1 \leq i \leq N)$. This more compact notation will be used in the remaining of this contribution.
 
In what follows, $\ket{S_j : \textbf{s}_i}$ and $\ket{S_j:A}$ represent the states in which the system $S_j$ is characterized by the $\textbf{s}_i$ spatial-spin-coordinate values in the first case and by the value $A$ of a property in the second case. We can also write $\ket{S_i : k_p}$, the state in which the $S_i$ system is in the $\ket{k_p}$ state and is described by the $k_p$ wavefunction. We have
$$
\braket{S_i : \textbf{s}_i| S_i : k_p} = \braket{\textbf{s}_i | k_p} = k_p(\textbf{s}_i),
$$
and an application in the space-spin representation of the closure relationship in the one-particle-state space
\begin{equation}\label{eq:ketIntket}
\ket{k_p} = \int _{S_4} \mathrm{d}\textbf{s}_0 \ket{\textbf{s}_0}\braket{\textbf{s}_0|k_p} = \int_{S_4} \mathrm{d}\textbf{s}_0 \ket{\textbf{s}_0}k_p(\textbf{s}_0). 
\end{equation}
This is obviously also true beyond the case of one-particle states. Note that in this contribution, integration on $S_4$ is understood as the combination of an integration on $\mathbb{R}^3$ and a sum on the spin-projection values (up ``$\uparrow$'' or down ``$\downarrow$'') since the one-particle states are defined on $S_4$, which is the Cartesian product between a continuous basis and a discrete basis. This consideration is also true below when we perform summation-integration on many-body states.
 
According to \eqref{eq:ketIntket}, the two $\hat{\lambda}_N^{\ell \rightarrow m}$ operators can be rewritten
\begin{align}\label{eq:lambdaNoperator}
\hspace*{-0.3cm} \hat{\lambda}_N^{\ell\rightarrow m}  &=  \int_{S_4} \!\!\!\mathrm{d}\textbf{s}_i \int_{S_{-4}} \!\!\!\!\!\!\!\mathrm{d}\textbf{s}_{-i} \int _{S_4}\!\!\! \mathrm{d}\textbf{s}_i' \int_{S_{-4}}\!\!\!\!\!\! \mathrm{d}\textbf{s}_{-i}'\,  \hat{O}_\lambda(\textbf{s}_i,\textbf{s}_{-i};\textbf{s}_i',\textbf{s}_{-i}')
\end{align}
with the $\hat{O}_\lambda$ operator expression being
\begin{align*}
\hat{O}_\lambda(\textbf{s}_i,\textbf{s}_{-i};\textbf{s}_i',\textbf{s}_{-i}') &= \psi ^\lambda_{\ell\rightarrow m} (\textbf{s}_i,\textbf{s}_{-i};\textbf{s}_i',\textbf{s}_{-i}') \\ & \times \ket{S_i : \textbf{s}_i ; S_{I\textbackslash \left\lbrace i \right\rbrace} : \textbf{s}_{-i}}\bra{S_i : \textbf{s}_i' ; S_{I\textbackslash \left\lbrace i \right\rbrace} : \textbf{s}_{-i}'}.
\end{align*}
In \eqref{eq:lambdaNoperator}, two integrations are performed over $S_4$, and two are performed over $S_{-4}$, which is defined as
$$S_{-4} \coloneqq \left[\mathbb{R}^3\times\left\lbrace \uparrow , \downarrow\right\rbrace\right]^{(N-1)}.$$
In the expression of $\hat{O}_\lambda$ given above, we see that if we perform the $(\lambda = \Delta)$ substitution, we find 
\begin{align}\nonumber
\psi ^\Delta_{\ell\rightarrow m} (\textbf{s}_i,\textbf{s}_{-i};\textbf{s}_i',\textbf{s}_{-i}') &= \psi_m (\textbf{s}_i,\textbf{s}_{-i}) \psi_m ^* (\textbf{s}_i',\textbf{s}_{-i}') \\ &- \psi_\ell (\textbf{s}_i,\textbf{s}_{-i}) \psi_\ell ^* (\textbf{s}_i',\textbf{s}_{-i}'),\label{eq:PsiDelta_i}
\end{align}
and if we rather perform the $(\lambda = \mathrm{T})$ substitution, we obtain
\begin{align*}
\psi ^\mathrm{T}_{\ell\rightarrow m} (\textbf{s}_i,\textbf{s}_{-i};\textbf{s}_i',\textbf{s}_{-i}') &= \psi_m (\textbf{s}_i,\textbf{s}_{-i}) \psi_\ell ^* (\textbf{s}_i',\textbf{s}_{-i}').
\end{align*}
Note that the ket and the bra can be rewritten as Kronecker tensor products
\begin{align*}&\ket{S_i : \textbf{s}_i ; S_{I\textbackslash \left\lbrace i \right\rbrace} : \textbf{s}_{-i}} \coloneqq \ket{S_i : \textbf{s}_i}\otimes \ket{S_{I\textbackslash \left\lbrace i \right\rbrace} : \textbf{s}_{-i}},\\&
\bra{S_i : \textbf{s}_i' ; S_{I\textbackslash \left\lbrace i \right\rbrace} : \textbf{s}_{-i}'} \coloneqq \bra{S_i : \textbf{s}_i' } \otimes \bra{S_{I\textbackslash \left\lbrace i \right\rbrace} : \textbf{s}_{-i}'}.
\end{align*}
The scalar product of two Kronecker tensor products obey certain rules. For instance, the following scalar product
\begin{align*}
\left(\bra{S_i:A}\otimes \bra{S_{I\textbackslash \left\lbrace i \right\rbrace}:B}\right)\left(\ket{S_i:A'}\otimes \ket{S_{I\textbackslash \left\lbrace i \right\rbrace}:B'}\right),
\end{align*}
corresponds to
$$\braket{S_i : A|S_i : A'} \braket{S_{I\textbackslash \left\lbrace i \right\rbrace} : B|S_{I\textbackslash \left\lbrace i \right\rbrace} : B'}=  \braket{A|A'}\braket{B|B'}.$$
With this in hand, we will now provide rules for going from the one-electron reduced difference and transition density operators to their corresponding spatial-spin representation (the kernels) and finite-dimensional matrix representations in the $\textbf{K}$ basis.

The expression for the reduction of the $N$-electron operators to the individual-electron difference and transition density operators is obtained after tracing out electrons — except the $i^\mathrm{th}$ one — from the $N$-electron operators:
$$
\hat{\gamma}_{i,\ell\rightarrow m}^{\lambda} \coloneqq \mathrm{tr}_{ \mathcal{S}_{I\textbackslash \left\lbrace i\right\rbrace}} \left(\hat{\lambda}_N^{\ell\rightarrow m}  \right).
$$
 Any $g\times h$ element $\left(\mathrm{with}\, (g,h) \in C^2\right)$ of the matrix representation in the \textbf{K} basis of the $\hat{\gamma}_{i,\ell\rightarrow m}^{\lambda}$ operators, i.e.,
$$\bm{\gamma}_{i,\ell\rightarrow m}^\lambda \coloneqq \mathcal{M}\left( \hat{\gamma}_{i,\ell\rightarrow m}^{\lambda},\textbf{K}\right)  , $$ 
reads
$$\left(\bm{\gamma}_{i,\ell\rightarrow m}^\lambda \right)_{g,h}\coloneqq \braket{ S_i : k_g |\hat{\gamma}_{i,\ell\rightarrow m}^{\lambda} |S_i : k_h}.$$
The expression of the $g\times h$ matrix element above can be detailed: we obtain, from the definition of the partial trace of an operator,
\begin{widetext}
\begin{align*}
\left(\bm{\gamma}_{i,\ell\rightarrow m}^\lambda\right)_{g,h} &= \sum _{r = 1}^{W}  \left(\bra{S_i : k_g}\otimes \bra{S_{I\textbackslash \left\lbrace i \right\rbrace} : w_r}\right)\left.\hat{\lambda}_N^{\ell\rightarrow m}\right.\left(\ket{S_i : k_h}\otimes \ket{S_{I\textbackslash \left\lbrace i \right\rbrace} : w_r}\right).
\end{align*}

$\;$

\noindent In order to detail the expression of the matrix element, we will insert the expression of the full operator, $\hat{\lambda}_N^{\ell\rightarrow m}$, obtained in equation \eqref{eq:lambdaNoperator}. This provides us with the following expression:

\begin{align*}
\hspace*{-0.3cm}\left(\bm{\gamma}_{i,\ell\rightarrow m}^\lambda\right)_{g,h} = \sum _{r = 1}^{W} \int_{S_4}\!\!\! \mathrm{d}\textbf{s}_i \int_{S_{-4}}\!\!\!\!\! \mathrm{d}\textbf{s}_{-i} \int_{S_4}\!\!\! \mathrm{d}\textbf{s}_i' \int_{S_{-4}}\!\!\!\!\!\! \mathrm{d}\textbf{s}_{-i}' &\left[\left(\right.\! \bra{S_i : k_g}\otimes \bra{S_{I\textbackslash \left\lbrace i \right\rbrace} : w_r}\right)\ket{S_i : \textbf{s}_i}\otimes \ket{S_{I\textbackslash \left\lbrace i \right\rbrace} : \textbf{s}_{-i}}\\
&\times \psi ^\lambda_{\ell\rightarrow m} (\textbf{s}_i,\textbf{s}_{-i};\textbf{s}_i',\textbf{s}_{-i}') 
 \bra{S_i : \textbf{s}_i'}\otimes \bra{S_{I\textbackslash \left\lbrace i \right\rbrace} : \textbf{s}_{-i}'}\left(\ket{S_i : k_h}\otimes \ket{S_{I\textbackslash \left\lbrace i \right\rbrace} : w_r}\!\left.\right)\right] .
\end{align*}
\end{widetext}
Using the rules for the scalar product of two Kronecker tensor product states we have briefly discussed above in the general case, we find
\begin{align*}
\hspace*{-0.3cm}\left(\bm{\gamma}_{i,\ell\rightarrow m}^\lambda\right)_{g,h} =  \int_{S_4} \!\!\!\mathrm{d}\textbf{s}_i \!\int_{S_{-4}}\!\!\!\!\!\! \mathrm{d}\textbf{s}_{-i}\! \int_{S_4}\!\!\! \mathrm{d}\textbf{s}_i' \!\int_{S_{-4}}\!\!\!\!\!\! \mathrm{d}\textbf{s}_{-i}' \, Q_\lambda^\textbf{q}(\textbf{s}_i,\textbf{s}_{-i};\textbf{s}_i',\textbf{s}_{-i}')
\end{align*}
with $\textbf{q} = (g,h,\ell,m)$, and the expression for $Q^\textbf{q}_\lambda$:
\begin{align*}
Q^\textbf{q}_\lambda(\textbf{s}_i,\textbf{s}_{-i};\textbf{s}_i',\textbf{s}_{-i}') &= \sum _{r = 1}^{W} k_g^*(\textbf{s}_i)\,w_r^*(\textbf{s}_{-i}) \,k_h(\textbf{s}_i') \,w_r(\textbf{s}_{-i}')\\&\times \psi ^\lambda_{\ell\rightarrow m} (\textbf{s}_i,\textbf{s}_{-i};\textbf{s}_i',\textbf{s}_{-i}') .
\end{align*}
Recalling equation \eqref{eq:ketIntket} and that, in $\mathcal{K}$, we can use a discrete closure relationship using the elements of the $\textbf{K}$ basis to write
\begin{equation}\label{eq:discrete_closure}
\forall \textbf{s}_i \in S_4,\,\ket{\textbf{s}_i} = \sum _{j=1}^K\ket{k_j}\braket{k_j|\textbf{s}_i},
\end{equation}
reminding also that in the one-particle spatial-spin basis, we have the following relationship that holds
$$\forall (\textbf{s}_i,\textbf{s}_i')\in \left(S_4\right)^2,\,\braket{\textbf{s}_i'|\textbf{s}_i} = \delta(\textbf{s}_i'-\textbf{s}_i),$$
where $\delta(\cdot - \cdot)$ is Dirac's delta distribution, we find that according to \eqref{eq:discrete_closure},
\begin{align*}
\hspace*{-0.3cm}\delta(\textbf{s}_i'-\textbf{s}_i) & \!= \sum _{j=1}^K \braket{\textbf{s}_i'|k_j}\braket{k_j|\textbf{s}_i}\\
&\!= \sum _{j=1}^K \int _{S_4}\!\!\mathrm{d}\textbf{s}_p\int _{S_4}\!\!\mathrm{d}\textbf{s}_q \braket{\textbf{s}_i'|\textbf{s}_p}\braket{\textbf{s}_q|\textbf{s}_i}k_j(\textbf{s}_p)k_j^*(\textbf{s}_q).
\end{align*}
Simplifications are then straightforward according to what precedes, and we conclude that, under the same conditions,
\begin{equation}\label{eq:s0's0delta}
\delta(\textbf{s}_i'-\textbf{s}_i)  = \sum _{j=1}^K k_j(\textbf{s}_i')k_j^*(\textbf{s}_i).
\end{equation}
Similar considerations in $\mathcal{W}$ lead to
\begin{equation} 
\forall(\textbf{s}_{-i},\textbf{s}_{-i}')\in\left(S_{-4}\right)^2,\,\delta (\textbf{s}_{-i}' - \textbf{s}_{-i}) = \sum _{r=1}^{W} w_r^*(\textbf{s}_{-i})w_r(\textbf{s}_{-i}').
\end{equation}
Accordingly, the expression for $Q_\lambda^\textbf{q}$ we have met above can first be rearranged 
\begin{align*}
\quad Q_\lambda^\textbf{q}(\textbf{s}_i,\textbf{s}_{-i};\textbf{s}_i',\textbf{s}_{-i}') &= \left(\sum _{r = 1}^{W}w_r^*(\textbf{s}_{-i})w_r(\textbf{s}_{-i}')\right) \\ &\times k_g^*(\textbf{s}_i)\, k_h(\textbf{s}_i') \,\psi ^\lambda_{\ell\rightarrow m} (\textbf{s}_i,\textbf{s}_{-i};\textbf{s}_i',\textbf{s}_{-i}')  .
\end{align*}
and then transformed into
\begin{align*}
Q_\lambda^\textbf{q}(\textbf{s}_i,\textbf{s}_{-i};\textbf{s}_i',\textbf{s}_{-i}')  &= k_g^*(\textbf{s}_i)\, k_h(\textbf{s}_i') \,\psi ^\lambda_{\ell\rightarrow m} (\textbf{s}_i,\textbf{s}_{-i};\textbf{s}_i',\textbf{s}_{-i}') \\ &\times \delta (\textbf{s}_{-i}' - \textbf{s}_{-i}).
\end{align*}
We then find the following final expression for the matrix elements $[(g,h) \in C^2]$:
\begin{align*}
\left(\bm{\gamma}^\lambda _{i,\ell \rightarrow m}\right)_{g,h} = \int_{S_4} \mathrm{d}\textbf{s}_i  \int_{S_4} \mathrm{d}\textbf{s}_i' \, k_g^*(\textbf{s}_i)    \,  \gamma^\lambda_{i,\ell\rightarrow m}(\textbf{s}_i;\textbf{s}_i') \, k_h(\textbf{s}_i') 
\end{align*}
where $\gamma^\lambda_{i,\ell\rightarrow m}$ is obtained after partially tracing $\psi_{\ell \rightarrow m}^\lambda$ on the $(N-1)$-dimensional space:
\begin{align}\label{eq:partial_trace_integral}
\hspace*{-0.2cm}\gamma^\lambda_{i,\ell\rightarrow m}(\textbf{s}_i;\textbf{s}_i') = \int_{S_{-4}}\!\!\!\!\!\!\! \mathrm{d}\textbf{s}_{-i}\int_{S_{-4}}\!\!\!\!\!\!\! \mathrm{d}\textbf{s}_{-i}'\, \psi_{\ell \rightarrow m}^\lambda(\textbf{s}_i,\textbf{s}_{-i};\textbf{s}_i',\textbf{s}_{-i}')  \delta(\textbf{s}_{-i}\!-\!\textbf{s}_{-i}') 
\end{align}
Notice the bold Greek letter ($\bm{\gamma}$) for the matrix representations in the \textbf{K} basis, and the unbold Greek letter ($\gamma$) for what we call the \textit{kernels}.
We are now interested in the corresponding total one-electron reduced difference and transition density \textit{operators}:
\begin{equation}\label{eq:TotalOperator}
\hat{\gamma}^{\lambda}_{\ell \rightarrow m} = \sum _{i = 1}^N \hat{\gamma} _{i,\ell \rightarrow m}^{\lambda}.
\end{equation}
Therefore, we have that the elements of the matrix representation of these operators in the $\textbf{K}$ basis have the following expression: for every $(g,h)$ in $C^2$, we have
\begin{align*}
\braket{S_i : k_g|\hat{\gamma}^{\lambda}_{\ell \rightarrow m}|S_i : k_h} = \sum _{i = 1}^N \braket{S_i : k_g|\hat{\gamma} _{i,\ell \rightarrow m}^{\lambda}|S_i : k_h}.
\end{align*}
If we denote the total one-electron reduced difference and transition \textit{density matrices} in the $\textbf{K}$ basis by
$$\bm{\gamma}_{\ell\rightarrow m}^\lambda \coloneqq \mathcal{M}\left(\hat{\gamma}_{\ell\rightarrow m}^{\lambda},\textbf{K}\right)  , $$
we have
\begin{equation}\label{eq:TotalMatRepres}
\bm{\gamma}_{\ell\rightarrow m}^\lambda = \sum _{i = 1}^N \bm{\gamma}_{i,\ell\rightarrow m}^\lambda.
\end{equation}
However, according to the fact that the $N$ electrons of the molecular system are undistinguishable, we immediately find that the expression for the matrix representation of the one-electron reduced difference and transition density operators in the $\textbf{K}$ basis is strictly independent of the electrons numbering. In other words,
$$\forall(i,j) \in I^2,\, \bm{\gamma}^\lambda_{i,\ell\rightarrow m} =\bm{\gamma}^\lambda_{j,\ell\rightarrow m}.$$
We therefore conclude that all the terms in the sum in the right-hand side of \eqref{eq:TotalMatRepres} are equal, and
$$\bm{\gamma}^\lambda _{\ell \rightarrow m} = N \bm{\gamma}^\lambda _{1,\ell \rightarrow m}.$$
For the matrix elements, we find, for every $(g,h)$ in $C^2$,
\begin{align} \label{eq:RDMel}
\left(\bm{\gamma}^\lambda _{\ell \rightarrow m}\right)_{g,h} = \int_{S_4}\!\!\! \mathrm{d}\textbf{s}_1  \int_{S_4}\!\!\! \mathrm{d}\textbf{s}_1' \, k_g^*(\textbf{s}_1)    \,  \gamma^\lambda_{\ell \rightarrow m}(\textbf{s}_1;\textbf{s}_1') \, k_h(\textbf{s}_1'), 
\end{align}
which allows to introduce two important objects:
\begin{equation}\label{eq:TotalKernel}
\gamma^\lambda_{\ell\rightarrow m}(\textbf{s}_1;\textbf{s}_1') = N\int_{S_{-4}}\!\!\! \mathrm{d}\textbf{s}_{-1}\,\psi_{\ell \rightarrow m}^\lambda(\textbf{s}_1,\textbf{s}_{-1};\textbf{s}_1',\textbf{s}_{-1}),
\end{equation}
i.e., the total one-body reduced difference ($\lambda = \Delta$) and transition ($\lambda = \mathrm{T}$) density kernels. 

We have just derived the expression of the one-electron reduced difference and transition density matrix elements by integrating the product of the corresponding kernels with two spinorbitals. In what follows, we establish the expression of the kernel from the matrix elements: We can rewrite the kernels values using Dirac's delta distributions according to:
\begin{align*}
\gamma^\lambda_{\ell \rightarrow m}(\textbf{s}_1 ; \textbf{s}_1') &= \int_{S_4}\!\!\! \mathrm{d}\textbf{s}_p \int_{S_4}\!\!\! \mathrm{d}\textbf{s}_q \, \gamma^\lambda_{\ell \rightarrow m}(\textbf{s}_p;\textbf{s}_q) \\ &\times \delta(\textbf{s}_1 - \textbf{s}_p)\delta(\textbf{s}_q-\textbf{s}_1').
\end{align*}
Inserting twice relationship \eqref{eq:s0's0delta}, we get the following expressions for the kernels values:
\begin{align*}
\gamma^\lambda_{\ell \rightarrow m}(\textbf{s}_1 ; \textbf{s}_1') &= \int_{S_4}\!\!\! \mathrm{d}\textbf{s}_p \int_{S_4}\!\!\! \mathrm{d}\textbf{s}_q \, \gamma^\lambda_{\ell \rightarrow m}(\textbf{s}_p;\textbf{s}_q) \\ & \times  \sum _{r=1}^{K} k_r^*(\textbf{s}_p)k_r(\textbf{s}_1) \sum _{s=1}^{K} k_s(\textbf{s}_q)k_s^*(\textbf{s}_1') .
\end{align*}
Rearranging the sums and integrating gives the following result:
\begin{equation}\label{eq:gammaKerSum}
\gamma^\lambda_{\ell \rightarrow m}(\textbf{s}_1 ; \textbf{s}_1') = \sum _{r=1}^{K}\sum _{s=1}^{K} \left(\bm{\gamma}^\lambda_{\ell \rightarrow m}\right)_{r,s}  k_r(\textbf{s}_1) k_s^*(\textbf{s}_1').
\end{equation}
where we find for every $(r,s)$ in $C^2$ the expression for the matrix elements we met in \eqref{eq:RDMel}.

We deduce now that any one-body reduced difference or transition kernel value is an element of the matrix representation of $\hat{\gamma}^{\lambda}_{\ell\rightarrow m}$:
\begin{equation}\label{eq:gammaLambdaOpBraKet}
\hat{\gamma}^{\lambda}_{\ell\rightarrow m} = \sum _{r=1}^{K}\sum _{s=1}^{K} \left(\bm{\gamma}^\lambda_{\ell \rightarrow m}\right)_{r,s} \ket{k_r}\bra{k_s}
\end{equation}
 in the spatial-spin representation:
$$\forall (\textbf{s}_1,\textbf{s}_1')\in \left(S_4\right)^2,\,\gamma^\lambda_{\ell \rightarrow m}(\textbf{s}_1 ; \textbf{s}_1') = \braket{\textbf{s}_1|\hat{\gamma}^{\lambda}_{\ell\rightarrow m}|\textbf{s}_1'}$$
and that any one-body reduced difference or transition density matrix element is an element of the matrix representation of $\hat{\gamma}^{\lambda}_{\ell\rightarrow m}$ in the $\textbf{K}$ basis:
$$\forall (p,q)\in C^2,\, \left(\bm{\gamma}^\lambda_{\ell\rightarrow m}\right)_{p,q} = \braket{k_p|\hat{\gamma}^{\lambda}_{\ell\rightarrow m}|k_q}.$$
Writing $(\textbf{s}_1) = (\textbf{r}_1,\sigma_1)$ and $(\textbf{s}_1') = (\textbf{r}_1',\sigma_1')$, we can trace out the spin from the kernel 
$$\gamma^{\lambda,\textbf{r}}_{\ell \rightarrow m}(\textbf{r}_1 ; \textbf{r}_1') = \sum _{\sigma_1 \in \left\lbrace \uparrow , \downarrow \right\rbrace}\sum _{\sigma_1' \in \left\lbrace \uparrow, \downarrow\right\rbrace} \gamma^\lambda_{\ell \rightarrow m}(\textbf{r}_1, \sigma_1 ; \textbf{r}_1',\sigma_1')\delta_{{\sigma _1^{\textcolor{white}{'}}} , {\sigma_1 '}}.$$
The one-electron reduced difference and transition \textit{spatial density functions} are then defined with the following expression:
\begin{align}\nonumber
n^{\lambda,\textbf{r}}_{\ell \rightarrow m}(\textbf{r}_1) &\coloneqq \int _{\mathbb{R}^3}\!\!\!\mathrm{d}\textbf{r}_1' \,\delta(\textbf{r}_1-\textbf{r}_1')\,\gamma^{\lambda,\textbf{r}}_{\ell \rightarrow m}(\textbf{r}_1 ; \textbf{r}_1') \\&= \gamma^{\lambda,\textbf{r}}_{\ell \rightarrow m}(\textbf{r}_1 ; \textbf{r}_1). \label{eq:TotalDens}
\end{align}
Note that setting simultaneously $(\lambda = \mathrm{T})$ and $(\ell = m)$ in \eqref{eq:TotalOperator}, in \eqref{eq:TotalMatRepres}, in \eqref{eq:TotalKernel}, and in \eqref{eq:TotalDens} gives the \textit{state} one-electron reduced density operator, matrix, kernel, and spatial density function, respectively.

$\;$

\noindent \textbf{Proposition III.A.1} \textit{All the} $\bm{\gamma}^\Delta_{\ell \rightarrow m}$ \textit{matrices} \textit{are Hermitian.}

$\;$

\noindent \textbf{Proof.} We start by proving that
$$\hspace*{-0.2cm}\forall (\ell,m)\in S^2,\forall (\textbf{s}_1,\textbf{s}_1')\in\left(S_4\right)^2,\, \gamma_{\ell\rightarrow m}^\Delta(\textbf{s}_1;\textbf{s}_1') = \left[\gamma_{\ell\rightarrow m}^\Delta(\textbf{s}_1';\textbf{s}_1)\right]^*.$$
For this sake, we first notice that if we set $(\lambda = \Delta)$, $(i=1)$, and $(\textbf{s}_{-1}=\textbf{s}_{-1}')$ in \eqref{eq:PsiDelta_i},
$$\psi^\Delta_{\ell\rightarrow m}(\textbf{s}_1,\textbf{s}_{-1};\textbf{s}_1',\textbf{s}_{-1}) = \left[\psi^\Delta_{\ell\rightarrow m}(\textbf{s}_1',\textbf{s}_{-1};\textbf{s}_1,\textbf{s}_{-1})\right]^*.$$
Reporting this in \eqref{eq:TotalKernel} leads to kernel values identity above. Hence, for every $(p,q)$ in $C^2$, we have
$$\hspace*{-0.3cm}\int_{S_4}\!\!\!\mathrm{d}\textbf{s}_1\!\int_{S_4}\!\!\!\mathrm{d}\textbf{s}_1' \,k_p^*(\textbf{s}_1) \left(\gamma^\Delta_{\ell \rightarrow m}(\textbf{s}_1;\textbf{s}_1') - \left[\gamma^\Delta_{\ell \rightarrow m}(\textbf{s}_1';\textbf{s}_1)\right]^*\right)k_q(\textbf{s}_1') = 0. $$
Using \eqref{eq:gammaKerSum} we can rewrite $\left[\gamma^\Delta_{\ell\rightarrow m}(\textbf{s}_1';\textbf{s}_1)\right]^*$ as
\begin{align*}\nonumber
\left[\gamma^\Delta_{\ell \rightarrow m}(\textbf{s}_1' ; \textbf{s}_1)\right]^* &= \sum _{r=1}^{K}\sum _{s=1}^{K} \left(\bm{\gamma}^\Delta_{\ell \rightarrow m}\right)^*_{r,s}  k_s(\textbf{s}_1)k_r^*(\textbf{s}_1') \\
&= \sum _{r=1}^{K}\sum _{s=1}^{K} \left[\left(\bm{\gamma}^\Delta_{\ell \rightarrow m}\right)^\dag\right]_{s,r}  k_s(\textbf{s}_1)k_r^*(\textbf{s}_1')
\end{align*}
and we deduce for the matrix elements, the equality
\begin{align*}
\left(\bm{\gamma}^\Delta_{\ell\rightarrow m}\right)_{p,q} &= \sum_{r=1}^K\sum_{s=1}^K  \int_{S_4}\!\!\!\mathrm{d}\textbf{s}_1\!\int_{S_4}\!\!\!\mathrm{d}\textbf{s}_1' \,\left[k_p^*(\textbf{s}_1) k_q(\textbf{s}_1')\right. \\
&\times \left(\gamma^\Delta_{\ell\rightarrow m}\right)_{r,s} \left.k_r(\textbf{s}_1)k_s^*(\textbf{s}_1') \right],
\end{align*}
is equivalent to
\begin{align*}
\left(\bm{\gamma}^\Delta_{\ell\rightarrow m}\right)_{p,q} & = \sum_{r=1}^K\sum_{s=1}^K  \int_{S_4}\!\!\!\mathrm{d}\textbf{s}_1\!\int_{S_4}\!\!\!\mathrm{d}\textbf{s}_1' \,\left[k_p^*(\textbf{s}_1) k_q(\textbf{s}_1')\right. \\
&\times \left[\left(\gamma^\Delta_{\ell\rightarrow m}\right)^\dag\right]_{s,r}  \left.k_s(\textbf{s}_1)k_r^*(\textbf{s}_1') \right].
\end{align*}
We therefore conclude, since $\textbf{K}$ is orthonormal, that
\begin{align*}
\left(\bm{\gamma}^\Delta_{\ell\rightarrow m}\right)_{p,q}=  \left[\left(\gamma^\Delta_{\ell\rightarrow m}\right)^\dag\right]_{p,q}.
\end{align*}
Since this is true for every $(p,q)$ in $C^2$, we complete the proof:
$$ \forall(\ell,m)\in S^2,\,\mathcal{M}\left( \hat{\gamma}^\Delta_{\ell \rightarrow m},\textbf{K}\right) \in \mathrm{Herm}_K$$
where $\mathrm{Herm}_K$ is used to denote the set of every $K$-dimensional Hermitian matrix. \\ \rightline{$\square$}
Note that this is also true for the state one-electron reduced density matrices. On the other hand, we cannot state that the one-electron reduced transition density matrices $\bm{\gamma}^\mathrm{T}_{\ell \rightarrow m}$ satisfy this condition when $\ell$ is different from $m$.
\subsection{Second quantization}
\noindent Since second quantization is often used in the literature in this context, we are going to derive the one-electron reduced difference and transition operators, matrices and kernels using a second quantization formulation of our problem. A short comment will be also given relatively to the notations used in the literature.

When acting in $\mathcal{S}$, a product of second quantization operators such as $\hat{k}_p^\dag\hat{k}_q^{\textcolor{white}{\dag}}$ reads
\begin{align*}
\hat{k}_p^\dag\hat{k}_q^{\textcolor{white}{\dag}} &\coloneqq \sum _{i\in I} \left(\ket{S_i : k_p}\bra{S_i : k_q}\otimes \hat{\mathbbm{1}}_{\mathcal{S}_{I\textbackslash\left\lbrace i\right\rbrace}}\right)
\end{align*}
with $\hat{k}_p^\dag$ creating an electron in state $\ket{k_p}$ and $\hat{k}_q$ annihilating an electron in state $\ket{k_q}$, and
\begin{align*}
\hat{\mathbbm{1}}_{\mathcal{S}_{I\textbackslash\left\lbrace i\right\rbrace}} \coloneqq \sum _{j=1}^W \ket{S_{I\textbackslash \left\lbrace{i}\right\rbrace} : w_j }\bra{S_{I\textbackslash \left\lbrace{i}\right\rbrace} : w_j }.
\end{align*}
We have basically introduced the one-particle excitation operator in second quantization in an orthonormal basis. Note that the orthonormality of the basis is a crucial parameter that we account for during all this section of the paper.

Let $\hat{\Gamma}(\textbf{s}_1;\textbf{s}_1')$ denote
$$\sum _{i\in I} \left(\ket{S_i : \textbf{s}_1'}\bra{S_i : \textbf{s}_1} \otimes \hat{\mathbbm{1}}_{\mathcal{S}_{I\textbackslash\left\lbrace i\right\rbrace}} \right).$$
Inserting twice the closure relation \eqref{eq:ketIntket} allows to rewrite it as
$$\sum _{i\in I}\sum_{p=1}^K\sum_{q=1}^K \left(k_p^*(\textbf{s}_1')k_q(\textbf{s}_1)\ket{S_i : k_p}\bra{S_i : k_q} \otimes \hat{\mathbbm{1}}_{\mathcal{S}_{I\textbackslash\left\lbrace i\right\rbrace}} \right),$$
which leads to the conclusion that
\begin{equation}\label{eq:FieldOpSQ}
\hat{\Gamma}(\textbf{s}_1;\textbf{s}_1') =\sum_{p=1}^K\sum_{q=1}^K k_p^*(\textbf{s}_1')k_q(\textbf{s}_1) \,\hat{k}_p^\dag\hat{k}_q^{\textcolor{white}{\dag}}. 
\end{equation}
This operator will be of seminal importance in the following derivations: we are going to use its matrix representation in the $\textbf{S}$ basis to derive the one-electron reduced difference and transition density matrices. 

Before that, we introduce two second-quantization operators, the so-called \textit{field operators}: 
$$\hat{\Psi}^\dag(\textbf{s}_1') = \sum _{p=1}^K k_p^*(\textbf{s}_1') \,\hat{k}_p^\dag$$
which creates an electron in state $\ket{\textbf{s}_1'}$, and
$$\hat{\Psi}(\textbf{s}_1) = \sum _{q=1}^K k_q(\textbf{s}_1) \,\hat{k}_q$$
which annihilates an electron in state $\ket{\textbf{s}_1}.$ It immediately comes that
$$
\hat{\Gamma}(\textbf{s}_1;\textbf{s}_1') = \hat{\Psi}^\dag(\textbf{s}_1') \hat{\Psi}(\textbf{s}_1).
$$
This operator is called the time-independent \textit{two-point field correlation operator}. 

$\;$

\noindent \textbf{Proposition III.B.1} For every $(\ell, m)$ in $S^2$, we have
$$\forall(\textbf{s}_1,\textbf{s}_1')\in\left(S_4\right)^2,\,\left\langle\!\psi _\ell\! \left| \hat{\Gamma}(\textbf{s}_1;\textbf{s}_1')\right|\!\psi _m\!\right\rangle = \gamma ^{\mathrm{T}}_{\ell \rightarrow m}(\textbf{s}_1;\textbf{s}_1').$$
\textbf{Proof.} We first insert the definition of $\hat{\Gamma}(\textbf{s}_1;\textbf{s}_1')$:
$$\left\langle\!\psi _\ell\! \left| \hat{\Gamma}(\textbf{s}_1;\textbf{s}_1')\right|\!\psi _m\!\right\rangle = \sum _{i\in I} \bra{ \psi_\ell} \left(\ket{S_i:\textbf{s}_1'}\bra{S_i:\textbf{s}_1}\otimes \hat{\mathbbm{1}}_{\mathcal{S}_{I\textbackslash\left\lbrace i\right\rbrace}}\right)\ket{\psi_m}$$
which transforms into
\begin{widetext}
\begin{align*}
\left\langle\!\psi _\ell\! \left| \hat{\Gamma}(\textbf{s}_1;\textbf{s}_1')\right|\!\psi _m\!\right\rangle 
&=  \sum _{i = 1}^{N}\sum _{r = 1}^{W} \int_{S_4}\!\!\! \mathrm{d}\textbf{s}_i \int_{S_{-4}}\!\!\!\!\! \mathrm{d}\textbf{s}_{-i} \int_{S_4}\!\!\! \mathrm{d}\textbf{s}_i' \int_{S_{-4}}\!\!\!\!\!\! \mathrm{d}\textbf{s}_{-i}' \,\psi^*_\ell(\textbf{s}_i';\textbf{s}_{-i}')\psi_m(\textbf{s}_i;\textbf{s}_{-i}) \\
&\times
 \bra{S_i : \textbf{s}_i'}\otimes \bra{S_{I\textbackslash \left\lbrace i \right\rbrace} : \textbf{s}_{-i}'} \left(\ket{S_i:\textbf{s}_1'}\bra{S_i:\textbf{s}_1}\otimes \ket{S_{I\textbackslash \left\lbrace{i}\right\rbrace} : w_r }\bra{S_{I\textbackslash \left\lbrace{i}\right\rbrace} : w_r }\right)\ket{S_i : \textbf{s}_i}\otimes \ket{S_{I\textbackslash \left\lbrace i \right\rbrace} : \textbf{s}_{-i}} \\
 &=   \sum _{i = 1}^{N} \int_{S_4}\!\!\! \mathrm{d}\textbf{s}_i \int_{S_{-4}}\!\!\!\!\! \mathrm{d}\textbf{s}_{-i} \int_{S_4}\!\!\! \mathrm{d}\textbf{s}_i' \int_{S_{-4}}\!\!\!\!\!\! \mathrm{d}\textbf{s}_{-i}' \,\psi_{\ell\rightarrow m}^\mathrm{T}(\textbf{s}_i,\textbf{s}_{-i};\textbf{s}_i',\textbf{s}_{-i}') \delta(\textbf{s}_i'-\textbf{s}_1')\delta({\textbf{s}_i-\textbf{s}_1})\left[\sum _{r = 1}^{W} w_r^*(\textbf{s}_{-i})w_r(\textbf{s}_{-i}')\right]. 
\end{align*}
\end{widetext}
The closure relationship in the $\textbf{W}$ basis reduces the square brackets in the last line, which leads to
$$\left\langle\!\psi _\ell\! \left| \hat{\Gamma}(\textbf{s}_1;\textbf{s}_1')\right|\!\psi _m\!\right\rangle  = \sum _{i=1}^N \int_{S_{-4}}\!\!\!\!\!\mathrm{d}\textbf{s}_{-i} \,\psi_{\ell\rightarrow m}^\mathrm{T}(\textbf{s}_1,\textbf{s}_{-i};\textbf{s}_1',\textbf{s}_{-i}).$$
Notice that $(\textbf{s}_1,\textbf{s}_{-i})$ means that the $i^\mathrm{th}$ electron is described by coordinate $\textbf{s}_1$ and the remaining system is described by $\textbf{s}_{-i}$, so $\textbf{s}_1$ does not necessarily correspond to the coordinate of electron $1$.

Since only the diagonal elements for the coordinates of the $S_{I\textbackslash\left\lbrace i\right\rbrace}$ systems are considered, the partitioning of the system is the same when described by $\psi _\ell$ and by $\psi _m$; since electrons are indistinguishable, all terms in the sum above are equal, leading to
\begin{align*}\left\langle\!\psi _\ell\! \left| \hat{\Gamma}(\textbf{s}_1;\textbf{s}_1')\right|\!\psi _m\!\right\rangle  &= N \int_{S_{-4}}\!\!\!\!\!\mathrm{d}\textbf{s}_{-1} \,\psi_{\ell\rightarrow m}^\mathrm{T}(\textbf{s}_1,\textbf{s}_{-1};\textbf{s}_1',\textbf{s}_{-1}) \\
&= \gamma^\mathrm{T}_{\ell \rightarrow m}(\textbf{s}_1;\textbf{s}_1').
\end{align*}
\begin{flushright}
$\square$
\end{flushright}
If we now turn to the alternative definition of the time-independent two-point field correlation operator in \eqref{eq:FieldOpSQ}, the $\ell \times m$ element of its matrix representation in the $\textbf{S}$ basis reads
\begin{align*}
\gamma^\mathrm{T}_{\ell \rightarrow m}(\textbf{s}_1;\textbf{s}_1')  &= \sum_{p=1}^K\sum_{q=1}^K k_p^*(\textbf{s}_1')k_q(\textbf{s}_1) \braket{\psi _\ell |\hat{k}_p^\dag\hat{k}_q^{\textcolor{white}{\dag}}|\psi_m}.
\end{align*}
For every $(r,s)$ in $C^2$, we have, according to \eqref{eq:RDMel},
\begin{align}\nonumber
\left(\bm{\gamma}^\mathrm{T}_{\ell \rightarrow m}\right)_{r,s} &= \int_{S_4}\!\!\!\mathrm{d}\textbf{s}_1\int_{S_4}\!\!\!\mathrm{d}\textbf{s}_1' \, k_r^*(\textbf{s}_1)k_s(\textbf{s}_1') \gamma^\mathrm{T}_{\ell \rightarrow m}(\textbf{s}_1;\textbf{s}_1')  \\
&= \braket{\psi _\ell |\hat{k}_s^\dag\hat{k}_r^{\textcolor{white}{\dag}}|\psi_m}.\label{eq:gammaTSQ}
\end{align}
Setting $(\ell = m)$ there, we find the state density matrix elements. Therefore, knowing that for every $(\ell,m)$ in $S^2$ and for every $(r,s)$ in $C^2$ we have
$$\left(\bm{\gamma}^\Delta_{\ell \rightarrow m}\right)_{r,s} = \left(\bm{\gamma}^\mathrm{T}_{m \rightarrow m}\right)_{r,s} - \left(\bm{\gamma}^\mathrm{T}_{\ell \rightarrow \ell} \right)_{r,s} ,$$
we conclude that the one-electron reduced difference density matrix elements in second quantization read
\begin{equation}\label{eq:gammaDeltaSQ}
\left(\bm{\gamma}^\Delta_{\ell \rightarrow m}\right)_{r,s} = \braket{\psi _m |\hat{k}_s^\dag\hat{k}_r^{\textcolor{white}{\dag}}|\psi_m} - \braket{\psi _\ell |\hat{k}_s^\dag\hat{k}_r^{\textcolor{white}{\dag}}|\psi_\ell}.
\end{equation}
The second-quantized expression of the matrix elements in \eqref{eq:gammaTSQ} and \eqref{eq:gammaDeltaSQ} can be directly introduced in \eqref{eq:gammaLambdaOpBraKet} to get the second-quantized expression of the one-electron reduced transition and difference density operators, respectively. We also have that \eqref{eq:gammaDeltaSQ} can be introduced in \eqref{eq:gammaKerSum} to get the $(\ell \rightarrow m)$ one-electron reduced difference density kernel, which can also be written as
$$\gamma^\Delta_{\ell \rightarrow m} (\textbf{s}_1;\textbf{s}_1') = \left\langle\! \psi _m \! \left| \hat{\Gamma}(\textbf{s}_1;\textbf{s}_1')\right|\!\psi_m \!\right\rangle- \left\langle\! \psi _\ell \! \left| \hat{\Gamma}(\textbf{s}_1;\textbf{s}_1')\right|\!\psi_\ell \!\right\rangle. $$
Finally, note that a very common practice is to consider that kind of notation:
$$\left(\bm{\gamma}^\mathrm{T}_{\ell m}\right)_{r,s} = \braket{\psi_\ell | \hat{k}_r^\dag\hat{k}_s^{\textcolor{white}{\dag}}|\psi_m},$$
and accordingly for the state $(\ell = m)$ one-electron reduced density matrices, and for the one-electron reduced difference density matrices. The reason why this is a very common practice is explained hereafter: in our developments, if one considers 
$$\left[\gamma_{\ell \rightarrow m}^\mathrm{T}(\textbf{s}_1';\textbf{s}_1)\right]^* = \gamma^\mathrm{T}_{m\rightarrow \ell}(\textbf{s}_1;\textbf{s}_1')$$
rather than $\gamma^\mathrm{T}_{\ell \rightarrow m}$, this leads, for every $(r,s)$ in $C^2$, to
$$\left(\bm{\gamma}^\mathrm{T}_{m\rightarrow \ell}\right)_{r,s} = \braket{\psi _m | \hat{k}_s^\dag\hat{k}_r^{\textcolor{white}{\dag}}|\psi_\ell} = \braket{\psi_\ell |\hat{k}^\dag_r\hat{k}_s^{\textcolor{white}{\dag}}|\psi_m}^*.$$
Hence,
$$\forall(\ell,m)\in S^2, \, \left(\bm{\gamma}^\mathrm{T}_{\ell \rightarrow m}\right)^\dag = \bm{\gamma}^\mathrm{T}_{m \rightarrow \ell},$$
or, consistent with our introductory remarks,
$$\forall(\ell,m)\in S^2, \, \left(\bm{\gamma}^\mathrm{T}_{\ket{\psi_\ell} \rightarrow \ket{\psi_m}}\right)^\dag = \bm{\gamma}^\mathrm{T}_{\bra{\psi_\ell} \rightarrow \bra{\psi_m}}.$$
When working with real-valued wavefunctions, the one-body transition properties values are equal for the ($\ell \rightarrow m$) and $(m \rightarrow \ell)$ transitions — i.e., for the $\ell \rightarrow m$ transition in $\mathcal{S}$ and in $\mathcal{S^*}$ — when the operator involved in the expectation value has a symmetric matrix representation in the one-body-state space. This excludes for instance odd powers of differential operators — as in the expression of the first-order non-adiabatic couplings for instance —, for they are anti-Hermitian. The most commonly studied quantity satisfying this condition is the electronic transition dipole moment, for which the corresponding operator is local and multiplicative. 

In fact, there has been a true notational ambiguity for decades: the two mechanics — in $\mathcal{S}$ and in $\mathcal{S}^*$  — were originally independently introduced in the 50's to derive the same quantities, and the two conventions have lived together until now. The effects of this ambiguity vanish in the application and use of the objects related in the present paper.

\noindent For the one-electron reduced state and difference density matrices, the reason is more simple: as we have seen before, those matrices are Hermitian (symmetric if the basis is real-valued). Consider the one-electron reduced density matrix corresponding to state $\ket{\psi _m}$ (with $m\in S$). Its elements read
$$\forall(r,s)\in C^2,\,\left(\bm{\gamma}^\mathrm{T}_{m\rightarrow m}\right)_{r,s} = \braket{\psi_m|\hat{k}_s^\dag\hat{k}_r^{\textcolor{white}{\dag}}|\psi_m}.$$
However,
$$\bm{\gamma}^\mathrm{T}_{m\rightarrow m} = \left(\bm{\gamma}^\mathrm{T}_{m\rightarrow m}\right)^\dag.$$
Hence, for every $(r,s)$ in $C^2$,
\begin{align*}
\left(\bm{\gamma}^\mathrm{T}_{m\rightarrow m}\right)_{r,s} =\left(\bm{\gamma}^\mathrm{T}_{m\rightarrow m}\right)_{s,r}^* = \braket{\psi_m|\hat{k}_r^\dag\hat{k}_s^{\textcolor{white}{\dag}}|\psi_m}^*.
\end{align*}
Since most of the practice is done with real-valued wavefunctions, it is very common to drop the star in the equalities above. This consideration naturally extends to the case of the one-electron reduced difference density matrices.
\section{Natural-orbital representation of molecular electronic transitions}\label{sec:NOs}
\noindent In this section we will discuss two types of natural-orbital bases: those in which the matrix representation of the one-electron difference and transition density operators is diagonal. Those are respectively called \textit{natural difference orbitals} and \textit{natural transition orbitals}. The former are used for defining the so-called detachment and attachment density matrices, while the latter are used for constructing the so-called transition-hole and transition-electron density matrices. The interpretation and discussion of these constructions will be reported in a further section.
\subsubsection{Natural difference orbitals}
\noindent Again, let $(\ell ,m) \in S^2$ be a couple of indices corresponding to the departure $(\ket{\psi_\ell})$ and the arrival $(\ket{\psi_m})$ electronic quantum states. The $\bm{\gamma}^{{\Delta}}_{\ell \rightarrow m}$ matrix is a $K\times K$ Hermitian matrix. Let 
\begin{equation*}
\textbf{U}^{\ell \rightarrow m} \coloneqq \left(\textbf{u}^{\ell \rightarrow m}_r\right)_{r\in C}
\end{equation*}
be the $K$--tuple of its eigenvectors, the ``natural \textit{difference} orbitals'' — their components in the $\textbf{K}$ basis are stored in the columns of $\textbf{U}^{\ell \rightarrow m} $ — and let
\begin{equation*}
\textbf{u}^{\ell \rightarrow m} \coloneqq \left(u^{\ell \rightarrow m}_r\right)_{r\in C}
\end{equation*}
be the $K$--tuple of its eigenvalues, the ``one-body-states transition occupation numbers''. From the natural difference orbitals and the absolute value of the negative occupation numbers, we then build the so-called one-electron detachment density matrix
\begin{equation*}
\hspace*{-0.1cm}\bm{\gamma}^d_{{\ell \rightarrow m}} \coloneqq -\,\textbf{U} ^{\ell \rightarrow m}\left[ \mathrm{diag}\left(\mathrm{min}\left(u^{\ell \rightarrow m}_r,0\right)\right)_{r\in C}\right]\left(\textbf{U}^{\ell \rightarrow m}\right)^\dag 
\end{equation*}
and from natural difference orbitals and the positive occupation numbers, we build the one-electron attachment density matrix 
\begin{equation*}
\bm{\gamma}^a_{{\ell \rightarrow m}} \coloneqq \textbf{U} ^{\ell \rightarrow m}\left[ \mathrm{diag}\left(\mathrm{max}\left(u^{\ell \rightarrow m}_r,0\right)\right)_{r\in C}\right]\left(\textbf{U}^{\ell \rightarrow m}\right)^\dag .
\end{equation*}
From what precedes, we see that the one-body reduced difference density operator can be written as a sum of $K$ dyads in the natural difference orbitals basis instead of a sum of $K^2$ dyads in the \textbf{K} basis:
\begin{align*}
\hat{\gamma}^\Delta_{{\ell \rightarrow m}} =  \sum _{r=1}^K u_r^{\ell \rightarrow m} \ket{u_r^{\ell \rightarrow m}}\bra{u_r^{\ell \rightarrow m}}
\end{align*}
where, for every $r$ in $C$ we have
\begin{align*}
\ket{u_r^{\ell \rightarrow m}} &= \sum _{s=1}^K \left(\textbf{U}^{\ell \rightarrow m}\right)_{s,r} \ket{k _s} = \sum _{s=1}^K \left(\textbf{u}_r^{\ell \rightarrow m}\right)_{s} \ket{k _s},\\
\bra{u_r^{\ell \rightarrow m}} &= \sum _{s=1}^K \left(\textbf{U}^{\ell \rightarrow m}\right)^*_{s,r} \bra{k _s} = \sum _{s=1}^K \left(\textbf{u}_r^{\ell \rightarrow m}\right)^*_{s} \bra{k _s}.
\end{align*}
Such a qualitative approach is valid for any excited-state calculation method, provided we have a ground- and an excited-state 1--RDM written in the same basis.
\subsubsection{Natural transition orbitals}
Before going into further details, we would like to recall two basic results from linear algebra: From the existence of a singular value decomposition for any finite-dimensional, complex-valued matrix, we get
\begin{lemma}\label{theo:SVDvalues}
Let ${\normalfont \textbf{A}}$ be an $m\times n$ complex matrix and ${\normalfont \textbf{A}}^\dag$ its adjoint. Let $p$ be any strictly positive integer lower or equal to $\mathrm{min}\left(m,n\right)$. Then, the $p^\mathrm{th}$ left-singular vector $\normalfont{\textbf{w}_p} \in \mathbb{C}^{m\times 1}$ of {\normalfont\textbf{A}} corresponds to the same singular value $\sigma _p$ as the $p^\mathrm{th}$ right-singular vector $\normalfont{\textbf{v}_p} \in \mathbb{C}^{n\times 1}$ of {\normalfont \textbf{A}}.
\end{lemma}
Lemma \ref{theo:SVDvalues} also reads
\begin{align}\label{eq:rsv}
\forall p \in \llbracket 1, \min (m,n)\rrbracket, \,&\textbf{A}\textbf{v}_p = \sigma_p \textbf{w}_p,
\\ \label{eq:lsv}
&\textbf{A}\!^\dag \textbf{w}_p = \sigma _p \textbf{v}_p.
\end{align}
\begin{lemma}\label{theo:eigenvalues}
Let ${\normalfont \textbf{A}}$ be an $m\times n$ complex matrix and ${\normalfont \textbf{A}}^\dag$ its adjoint. Let $q$ be the strictly positive integer equal to $\mathrm{min}\left(m,n\right)$. Then, ${\normalfont \textbf{AA}^\dag}$ and ${\normalfont\textbf{A}^\dag\textbf{A}}$ share $q$ eigenvalues. Those eigenvalues are the squared singular values of {$\normalfont \textbf{A}$}.
\end{lemma}
Proof: Let $p$ be any strictly positive integer lower or equal to $q$. Then, multiplying \eqref{eq:rsv} to the left by $\textbf{A}^\dag$ leads, according to \eqref{eq:lsv}, to
\begin{equation}\label{eq:squaredsvo}
\textbf{A}^\dag\textbf{A}\textbf{v}_p = \sigma_p^2 \textbf{v}_p,
\end{equation}
while multiplying \eqref{eq:lsv} to the left by \textbf{A} leads, according to \eqref{eq:rsv}, to
\begin{equation}\label{eq:squaredsvv}
\textbf{A}\textbf{A}^\dag \textbf{w}_p = \sigma _p^2 \textbf{w}_p.
\end{equation}
\begin{flushright}
$\square$
\end{flushright}
\noindent Any general $\bm{\gamma}^{{\mathrm{T}}}_{\ell \rightarrow m}$ is a $K \times K$ complex-valued matrix. Let 
\begin{equation*}
\textbf{L}^{\ell \rightarrow m} \coloneqq \left(\bm{l}^{\ell \rightarrow m}_p\right)_{p\in C}
\end{equation*}
be the $K$--tuple of its {left}-singular vectors, the so-called ``{left} natural \textit{transition} orbitals'' in the $\textbf{K}$ basis. Let 
\begin{equation*}
\textbf{R}^{\ell \rightarrow m} \coloneqq \left(\bm{r}^{\ell \rightarrow m}_p\right)_{p\in C}
\end{equation*}
be the $K$--tuple of its {right}-singular vectors, the so-called ``{right} natural \textit{transition} orbitals'' in the $\textbf{K}$ basis,  and let
\begin{equation*}
\bm{\lambda}^{\ell \rightarrow m} \coloneqq \left(\lambda^{\ell \rightarrow m}_r\right)_{r\in C}
\end{equation*}
be the $K$--tuple of its singular values. According to lemma \ref{theo:SVDvalues}, the left and right singular vectors are always paired (each pair shares a singular value). This is typical of the natural transition orbitals — the natural difference orbitals are, in general, unpaired. From what precedes, we get
\begin{equation*}
\hat{\gamma}^\mathrm{T}_{\ell \rightarrow m} =  \sum _{p=1}^K \lambda_p^{\ell \rightarrow m} \ket{l^{\ell \rightarrow m}_p}\bra{r^{\ell \rightarrow m}_p}  
\end{equation*}
where, for every $p$ in $C$ we have
\begin{align*}
\ket{l_p^{\ell \rightarrow m}} &= \sum _{s=1}^K \left(\textbf{L}^{\ell \rightarrow m}\right)_{s,p} \ket{k _s} = \sum _{s=1}^K \left(\bm{l}_p^{\ell \rightarrow m}\right)_{s} \ket{k _s},\\
\bra{r_p^{\ell \rightarrow m}} &= \sum _{s=1}^K \left(\textbf{R}^{\ell \rightarrow m}\right)^*_{s,p} \bra{k _s} = \sum _{s=1}^K \left(\bm{r}_p^{\ell \rightarrow m}\right)^*_{s} \bra{k _s}.
\end{align*}
Finally, consider the transition ``left'' operator
\begin{equation*}
\hat{\gamma}^l_{\ell \rightarrow m} = \sum _{p=1}^K \left(\lambda^{\ell \rightarrow m}_p\right)^2 \ket{l^{\ell \rightarrow m}_p}\bra{l^{\ell \rightarrow m}_p} ,
\end{equation*}
and its matrix representation in the \textbf{K} basis:
\begin{equation*}
\mathcal{M}\left(\hat{\gamma}^l_{\ell \rightarrow m} , \textbf{K}\right) = \bm{\gamma}_{\ell \rightarrow m}^\mathrm{T}\!\!\left(\bm{\gamma}_{\ell \rightarrow m}^\mathrm{T}\right)^{\dag}
\end{equation*}
that we will write $\bm{\gamma}^l_{\ell\rightarrow m}$. We also consider the transition ``right'' operator
\begin{equation*}
\hat{\gamma}^r_{\ell \rightarrow m} = \sum _{p=1}^K \left(\lambda^{\ell \rightarrow m}_p\right)^2 \ket{r^{\ell \rightarrow m}_p}\bra{r^{\ell \rightarrow m}_p} 
\end{equation*}
and matrix representation in the \textbf{K} basis:
\begin{equation*}
\mathcal{M}\left(\hat{\gamma}^r_{\ell \rightarrow m} , \textbf{K}\right) = \left(\bm{\gamma}_{\ell \rightarrow m}^\mathrm{T}\right)^{\dag}\bm{\gamma}_{\ell \rightarrow m}^\mathrm{T}
\end{equation*}
that we will write $\bm{\gamma}^r_{\ell\rightarrow m}$. From lemma \ref{theo:eigenvalues}, we have that the left-singular vectors of the one-body reduced transition density operator are the eigenvectors of the $l$-density matrix, while its right-singular vectors are the eigenvectors of the $r$-density matrix.

Those matrices are often involved in qualitative and quantitative analyses of molecular electronic transitions. Their physical content, and the motivation for their use is discussed hereafter.

\subsubsection{Electron-hole correlation function}

Consistent with what is done in the literature, in what follows, we consider the special case ($\ell = 0$), with $(\braket{\psi _0|\psi_0} =1)$, and we write ``$\mathrm{i}$'' the pure-imaginary number ($\mathrm{i}^2 = -1$).

Consider the time-dependent extension of the ground-state one-electron reduced density kernel, the one-fermion (hereafter denoted one-body) Green function \cite{strinati_effects_1984}
\begin{align*}
\mathrm{i}G_1(\textbf{s}_1,t_1;\textbf{s}_1',\textbf{t}_1') = \braket{\psi _0 | \hat{T}[\hat{\Psi}_{\mathrm{st}}(\textbf{s}_1,t_1) \hat{\Psi}_{\mathrm{st}}^\dag(\textbf{s}_1',t_1')]|\psi _0}
\end{align*}
where $\hat{T}[\,\cdot\,]$ is Wick's time-ordering operator which places the operator with increasing time from the right to the left. In what precedes, ``st'' stands for ``space-time''. The $\hat{\Psi}_{\mathrm{st}}$ and $\hat{\Psi}^\dag_{\mathrm{st}}$ operators are the time-dependent extension to spin-spatial field operators met in section \ref{sec:AppRDMs} (hence, the $\mathrm{st}$ subscript), which, in the Heisenberg picture, read
\begin{align*}
(\textbf{s},t) \longmapsto \hat{\Psi}_{\mathrm{st}} (\textbf{s},t) &= \mathrm{e}^{\mathrm{i}\hat{H}t} \hat{\Psi}(\textbf{s}) \mathrm{e}^{-\mathrm{i}\hat{H}t} ,\\
(\textbf{s},t) \longmapsto \hat{\Psi}_{\mathrm{st}}^\dag (\textbf{s},t) &= \mathrm{e}^{\mathrm{i}\hat{H}t} \hat{\Psi}^\dag(\textbf{s}) \mathrm{e}^{-\mathrm{i}\hat{H}t}.
\end{align*}
$\hat{H}$ is the reference Hamiltonian operator. For the sake of notational brevity, we will use 
$$(1) \coloneqq (\textbf{s}_1, t_1) ,  \, (1') \coloneqq (\textbf{s}_1',t_1'),$$
and 
$$\hat{\Psi}_1 \coloneqq \hat{\Psi}_{\mathrm{st}} (\textbf{s}_1,t_1), \;\hat{\Psi}_{1'}^\dag \coloneqq \hat{\Psi}_{\mathrm{st}}^\dag (\textbf{s}_1',t_1').$$ 
The one-body Green function alternatively reads
\begin{align*}
\mathrm{i}G_1(1;1') &= \Theta(t_1 - t_1') \braket{\psi _0 | \hat{\Psi}_1^{\textcolor{white}{\dag}} \hat{\Psi}^\dag_{1'}|\psi _0} \\
&- \Theta(t_1' - t_1) \braket{\psi _0 | \hat{\Psi}^\dag_{1'}\hat{\Psi}_1^{\textcolor{white}{\dag}} |\psi _0},
\end{align*}
where $\Theta$ is the Heaviside step function, i.e., the antiderivative of Dirac's delta distribution. It takes a value of $1$ when its argument is positive, and $0$ when its argument is negative. 

We now apply the operator rules to derive
\begin{align*}
\mathrm{i}G_1(1;1') &= \mathrm{i}G_{1,+}(1;1') - \mathrm{i}G_{1,-}(1;1'),
\end{align*}
with $\mathrm{i}G_{1,+}(1;1')$ being equal to
\begin{align*}
  \Theta(t_1 - t_1')\mathrm{e}^{\mathrm{i}E_0(t_1-t_1')} \braket{\psi _0 | \hat{\Psi} (\textbf{s}_1)\mathrm{e}^{-\mathrm{i}\hat{H}(t_1-t_1')}\hat{\Psi}^\dag(\textbf{s}_{1'}) |\psi _0},
\end{align*}
and $\mathrm{i}G_{1,-}(1;1')$ being equal to
$$\Theta(t_1' - t_1)\mathrm{e}^{-\mathrm{i}E_0(t_1-t_1')} \braket{\psi _0 | \hat{\Psi}^\dag (\textbf{s}_1')\mathrm{e}^{\mathrm{i}\hat{H}(t_1-t_1')}\hat{\Psi}(\textbf{s}_1) |\psi _0}.$$
In the two expressions above, $E_0$ is the ground electronic state energy. We immediately see the connection with the ground-state one-body reduced density kernel:
$$\gamma^\mathrm{T}_{0 \rightarrow 0} (\textbf{s}_1;\textbf{s}_1') = -\mathrm{i}G_1(\textbf{s}_1,t_1;\textbf{s}_1',t_1^+)$$
where $\displaystyle t_1^+ = \lim_{\delta t \rightarrow 0^+}(t_1 + \delta t)$.

The two-fermion (hereafter denoted two-body) Green function \cite{strinati_effects_1984} is similarly defined as
$$\mathrm{i}^2G_2(1,2 ; 1',2') = \braket{\psi _0 | \hat{T}[\hat{\Psi}(1)\hat{\Psi}(2)\hat{\Psi}^\dag(2')\hat{\Psi}^\dag(1') ]|\psi _0}.$$
Depending on the time ordering, it describes the correlated evolution of two particles, two holes, or a one-hole-one-particle pair.

\noindent The correlation between the evolution of two bodies is measured by the four-point correlation function:
$$-L(1,2;1',2') = \mathrm{i}G_2(1,2;1',2') - \mathrm{i}G_1(1;1')G_1(2;2').$$
This quantity is central in the framework of the so-called \textit{Bethe-Salpeter equation} for neutral excitations \cite{strinati_effects_1984}.

For every $x$ in $\left\lbrace 1,2\right\rbrace$, we set the notation $$(x^+) \coloneqq (\textbf{s}_x',t_x^+),$$ with
\begin{equation}\label{eq:lim0+}
  t_x^+\coloneqq \lim_{\delta t \rightarrow 0^+} \left(t_x +\delta t\right),
\end{equation}
Appropriate time ordering leads to the sum of two functions describing the coupled evolution of an electron and a hole:
\begin{align*}
\mathrm{i}^2G_2(1,2;1^+,2^+) &\coloneqq (-1)^{4}\mathrm{i}^2G^{\mathrm{ph}}_{2,+}(1,2;1^+,2^+) \\ &+ (-1)^4\mathrm{i}^2G^{\mathrm{ph}}_{2,-}(1,2;1^+,2^+),
\end{align*}
with, for the expression of $G^{\mathrm{ph}}_{2,+}$,
\begin{align*}
 \mathrm{i}^2G^{\mathrm{ph}}_{2,+}(1,2;1^+,2^+) = \Theta(t_1-t_2)\braket{\psi _0 |\hat{\Psi}_{1^+}^\dag\hat{\Psi}_{1}^{\textcolor{white}{\dag}}\hat{\Psi}_{2^+}^\dag\hat{\Psi}_{2}^{\textcolor{white}{\dag}}  |\psi _0},
\end{align*}
and, for the expression of $G^{\mathrm{ph}}_{2,-}$,
\begin{align*}
\mathrm{i}^2G^{\mathrm{ph}}_{2,-}(1,2;1^+,2^+) = \Theta(t_2-t_1)\braket{\psi _0 |\hat{\Psi}_{2^+}^\dag\hat{\Psi}_{2}^{\textcolor{white}{\dag}}\hat{\Psi}_{1^+}^\dag\hat{\Psi}_{1}^{\textcolor{white}{\dag}}  |\psi _0}.
\end{align*}
$G_2(1,2;1^+,2^+)$ describes the evolution (creation, propagation, destruction) of a bound one-hole-one-electron pair.

Therefore, choosing an appropriate time ordering allows to probe the causal four-point, two-time one-electron-one-hole (linear) response of the system:
\begin{align*}
 -L(1,2 ; 1^+,2^+) &= \mathrm{i} G_2(1,2;1^+,2^+)  \\&- \mathrm{i}G_1(1;1^+)G_1(2;2^+). 
\end{align*}
It compares two correlated single excitations (the one-hole-one-electron pair is created, then space-time propagated and finally annihilated) and two uncorrelated ones.

If $\textbf{S}$ is infinite-dimensional, we are allowed to use the closure relationship in the expression of $\mathrm{i}^2G^{\mathrm{ph}}_{2,+}(1,2;1^+,2^+)$, which turns it into
\begin{align*}
 \sum _{q=0}^\infty \Theta(t_1-t_2)\braket{\psi _0 |\hat{\Psi}_{1^+}^\dag\hat{\Psi}_{1}^{\textcolor{white}{\dag}}|\psi _q}\braket{\psi _q|\hat{\Psi}_{2^+}^\dag\hat{\Psi}_{2}^{\textcolor{white}{\dag}}  |\psi _0}.
\end{align*}
Note that for a time-independent Hamiltonian operator, both the one- and two-body Green functions are invariant upon collective coordinate translation. Setting $(\tau \coloneqq t_1-t_2)$, we find that the arguments of $G_2$, i.e. ($1,2;1^+,2^+$), are transformed into $(\textbf{s}_1,\textbf{s}_2;\textbf{s}_1',\textbf{s}_2';\tau)$. We also set
$$\omega_q \coloneqq E_q-E_0.$$
Expanding the field operators and recalling \eqref{eq:lim0+}, we obtain 
$$\hspace*{-0.3cm}\mathrm{i}^2G_{2,+}^\mathrm{ph}(\textbf{s}_1,\textbf{s}_2;\textbf{s}_1',\textbf{s}_2';\tau) = \Theta(\tau) \sum _{q=0}^\infty \gamma ^\mathrm{T}_{0\rightarrow q} (\textbf{s}_1;\textbf{s}_1')\gamma ^\mathrm{T}_{q\rightarrow 0} (\textbf{s}_2;\textbf{s}_2')\mathrm{e}^{-\mathrm{i} \omega _q \tau}.$$
For this, we used the results from section \ref{sec:AppRDMs}, i.e.,
\begin{align*}
\braket{\psi _0 |\hat{\Psi}^\dag(\textbf{s}_1')\hat{\Psi}(\textbf{s}_1)|\psi _q} &=\gamma ^\mathrm{T}_{0\rightarrow q} (\textbf{s}_1;\textbf{s}_1'),\\
\braket{\psi _q|\hat{\Psi}^\dag(\textbf{s}_2')\hat{\Psi}(\textbf{s}_2)  |\psi _0} &=\gamma ^\mathrm{T}_{q\rightarrow 0} (\textbf{s}_2;\textbf{s}_2').
\end{align*}
Similarly, we obtain that 
$$\hspace*{-0.4cm}\mathrm{i}^2G_{2,-}^\mathrm{ph}(\textbf{s}_1,\textbf{s}_2;\textbf{s}_1',\textbf{s}_2';\tau) = \Theta(-\tau) \sum _{q=0}^\infty \gamma ^\mathrm{T}_{0\rightarrow q} (\textbf{s}_2;\textbf{s}_2')\gamma ^\mathrm{T}_{q\rightarrow 0} (\textbf{s}_1;\textbf{s}_1')\mathrm{e}^{\mathrm{i} \omega _q \tau}.$$
Due to the definition of the four time-dependent variables in $L$, we have that $\mathrm{i}G_1(1;1^+)G_1(2;2^+)$ reads
\begin{align*}
\mathrm{i}G_1(1;1^+)G_1(2;2^+) &= -\mathrm{i} Q_0(\textbf{s}_1,\textbf{s}_2 ; \textbf{s}_1',\textbf{s}_2') \\
&= -\mathrm{i}\Theta(\tau) Q_0(\textbf{s}_1,\textbf{s}_2 ; \textbf{s}_1',\textbf{s}_2') \\
& -\mathrm{i}\Theta(-\tau) Q_0(\textbf{s}_1,\textbf{s}_2 ; \textbf{s}_1',\textbf{s}_2') ,
\end{align*}
where we have defined
\begin{align*}
Q_0(\textbf{s}_1,\textbf{s}_2 ; \textbf{s}_1',\textbf{s}_2') &\coloneqq \braket{\psi _0|\hat{\Psi}^\dag (\textbf{s}_1')\hat{\Psi}(\textbf{s}_1)|\psi _0}\\&\times\braket{\psi _0|\hat{\Psi}^\dag (\textbf{s}_2')\hat{\Psi}(\textbf{s}_2)|\psi _0} \\
&= \gamma ^\mathrm{T}_{0\rightarrow 0} (\textbf{s}_2;\textbf{s}_2')\gamma ^\mathrm{T}_{0\rightarrow 0} (\textbf{s}_1;\textbf{s}_1').
\end{align*}
%
We see that the ($q=0$) term in $\mathrm{i}^2G_2(\textbf{s}_1,\textbf{s}_2;\textbf{s}_1',\textbf{s}_2';\tau)$ reads
$$P_0 (\textbf{s}_1,\textbf{s}_2;\textbf{s}_1',\textbf{s}_2') \coloneqq \left[\Theta(\tau) + \Theta(-\tau)\right]Q_0(\textbf{s}_1,\textbf{s}_2;\textbf{s}_1',\textbf{s}_2').$$
Since $(-\mathrm{i})\mathrm{i}^2G_2  = \mathrm{i}G_2 $, we can write
$$-\mathrm{i}P_0 (\textbf{s}_1,\textbf{s}_2;\textbf{s}_1',\textbf{s}_2') = \mathrm{i}G_1(1;1^+)G_1(2;2^+),$$
so $\mathrm{i}G_1(1;1^+)G_1(2;2^+)$ and the ($q=0$) term of $\mathrm{i}^2G_2(\textbf{s}_1,\textbf{s}_2;\textbf{s}_1',\textbf{s}_2';\tau)$ cancel each other in $L$.
We now turn to the electron-hole correlation function, whose expression introduces a convergence factor to $L$:
\begin{align*}
-\chi (\textbf{s}_1,\textbf{s}_2;\textbf{s}_1',\textbf{s}_2';\tau) &\coloneqq \lim _{\eta ^+ \rightarrow 0^+} \mathrm{e}^{-\alpha \tau \eta^+}L(\textbf{s}_1,\textbf{s}_2;\textbf{s}_1',\textbf{s}_2';\tau)
\end{align*}
with $\eta^+ \in \mathbb{R}_+$, and $[\alpha = \mathrm{sgn}(\tau)].$
Since
\begin{align*}
&\forall \tau \in \mathbb{R}^*, \, \mathrm{e}^{-\alpha \tau \eta^+}\Theta(\tau) = \mathrm{e}^{-\tau \eta^+}\Theta(\tau),\\
&\forall \tau \in \mathbb{R}^*, \, \mathrm{e}^{-\alpha \tau \eta^+}\Theta(-\tau) = \mathrm{e}^{\tau \eta^+}\Theta(-\tau),
\end{align*}
the expression of $\chi$ can be rewritten
\begin{align*}
\chi (\textbf{s}_1,\textbf{s}_2;\textbf{s}_1',\textbf{s}_2';\tau)  &= \lim _{\eta ^+ \rightarrow 0^+}\left[G_{2,>}^\mathrm{ph}(\textbf{s}_1,\textbf{s}_2;\textbf{s}_1',\textbf{s}_2';\tau;\eta^+) \right.\\
&+ \left.G_{2,<}^\mathrm{ph}(\textbf{s}_1,\textbf{s}_2;\textbf{s}_1',\textbf{s}_2';\tau;\eta^+)\right]
\end{align*}
with its two components (one for $\tau <0$ and one for $\tau > 0$) being
\begin{align*}
G_{2,>}^\mathrm{ph}(\textbf{s}_1,\textbf{s}_2;\textbf{s}_1',\textbf{s}_2';\tau;\eta^+) &\coloneqq -\mathrm{i} \Theta(\tau)\mathrm{e}^{-\tau\eta ^+}\mathrm{e}^{-\mathrm{i} \omega _q \tau} \\ &\times \sum _{q=1}^\infty \gamma ^\mathrm{T}_{0\rightarrow q} (\textbf{s}_1;\textbf{s}_1')\gamma ^\mathrm{T}_{q\rightarrow 0} (\textbf{s}_2;\textbf{s}_2'), \\
G_{2,<}^\mathrm{ph}(\textbf{s}_1,\textbf{s}_2;\textbf{s}_1',\textbf{s}_2';\tau;\eta^+) &\coloneqq -\mathrm{i} \Theta(-\tau)\mathrm{e}^{\tau\eta ^+}\mathrm{e}^{\mathrm{i} \omega _q \tau} \\ &\times \sum _{q=1}^\infty \gamma ^\mathrm{T}_{0\rightarrow q} (\textbf{s}_2;\textbf{s}_2')\gamma ^\mathrm{T}_{q\rightarrow 0} (\textbf{s}_1;\textbf{s}_1').
\end{align*}
We will now examine in details how we can Fourier-transform the electron-hole correlation function. Let $g_>$ be the $[\tau \longmapsto e^{-\tau \eta ^+}\Theta(\tau)]$ map. Its Fourier transform reads
$$\hat{F}g_>(\omega) = \dfrac{1}{\eta^+ + \mathrm{i}\omega}$$
and we immediately find $-\mathrm{i} \hat{F}g_>(\omega) = \left(-\omega + \mathrm{i}\eta ^+\right)^{-1}$.

Given a map $y_1$ whose Fourier transform reads $[\omega \longmapsto \hat{F}y_1(\omega)]$, the Fourier transform of the time-scaled map $[t \longmapsto  y_2(t) = y_1(\alpha t)]$ with $\alpha$ real and different from zero, is
$$\hat{F}y_2 \, : \, \omega \longmapsto \hat{F}y_2(\omega) = \dfrac{\hat{F}y_1({\omega}/{\alpha})}{|\alpha |}.$$
We therefore deduce that the Fourier transform of $g_< \, : \, [\tau \longmapsto e^{\tau \eta ^+}\Theta(-\tau)]$ is
$$\hat{F}g_<(\omega) = \dfrac{1}{\eta^+ - \mathrm{i}\omega}$$
and we immediately find $-\mathrm{i} \hat{F}g_<(\omega) = \left(\omega +\mathrm{i}\eta ^+\right)^{-1}$.
We know that, given a map, say $h_1$, whose Fourier transform is $[\omega \longmapsto \hat{F}h_1(\omega)]$, the Fourier transform corresponding to the shifted map $[x \longmapsto  \mathrm{e}^{\mathrm{i}\omega_0t}h_1(x)]$ (with $\omega _0 \in \mathbb{R}$ in our application of interest) will simply be $$[\omega \longmapsto \hat{F}h_1(\omega - \omega _0)].$$ We conclude, setting
\begin{align*}
\Omega _{0 \rightarrow q} &\coloneqq \omega _q, \\
\Omega _{q \rightarrow 0} &\coloneqq - \omega _q,
\end{align*}
that
\begin{align*}
-\mathrm{i}\hat{F}g_>(\omega - (-\omega _q)) &= \left( \Omega _{q\rightarrow 0} - \omega + i\eta^+\right)^{-1},\\
-\mathrm{i}\hat{F}g_<(\omega - \omega _q) &= \left(\omega - \Omega _{0\rightarrow q} + i\eta^+\right)^{-1},
\end{align*}
and that the Fourier transform of $\chi $ is
\begin{align*}
\chi(\textbf{s}_1,\textbf{s}_2;\textbf{s}_1',\textbf{s}_2';\omega)   &= \chi ^<(\textbf{s}_1,\textbf{s}_2;\textbf{s}_1',\textbf{s}_2';\omega) \\&- \chi ^>(\textbf{s}_1,\textbf{s}_2;\textbf{s}_1',\textbf{s}_2';\omega),
\end{align*}
where $\chi ^>$ and $\chi ^<$ have the following expression:
\begin{align*}
\hspace*{-.3cm}\chi ^<(\textbf{s}_1,\textbf{s}_2;\textbf{s}_1',\textbf{s}_2';\omega) &= \lim _{\eta ^+ \rightarrow 0^+}\sum _{q=1}^\infty \dfrac{\gamma ^\mathrm{T}_{0\rightarrow q} (\textbf{s}_2;\textbf{s}_2')\gamma ^\mathrm{T}_{q\rightarrow 0} (\textbf{s}_1;\textbf{s}_1')}{  \omega - \Omega _{0\rightarrow q}  + i\eta^+},\\
\hspace*{-.3cm}\chi ^>(\textbf{s}_1,\textbf{s}_2;\textbf{s}_1',\textbf{s}_2';\omega) &= \lim _{\eta ^+ \rightarrow 0^+}\sum _{q=1}^\infty \dfrac{\gamma ^\mathrm{T}_{0\rightarrow q} (\textbf{s}_1;\textbf{s}_1')\gamma ^\mathrm{T}_{q\rightarrow 0} (\textbf{s}_2;\textbf{s}_2')}{ \omega - \Omega _{q\rightarrow 0} - i\eta^+}.
\end{align*}
We deduce that $\chi $ is meromorphic, with poles at the imaginary-shifted exact electronic transition energies. For a given $(q=m)$, the residue of the $m^\mathrm{th}$ term of $\chi ^<$, i.e., corresponding to the $(\Omega _{0\rightarrow m}- \mathrm{i}\eta ^+)$ pole, reads
\begin{align*}
R_m^<(\textbf{s}_1,\textbf{s}_2;\textbf{s}_1',\textbf{s}_2') &= \gamma ^\mathrm{T}_{0\rightarrow m} (\textbf{s}_2;\textbf{s}_2')\gamma ^\mathrm{T}_{m\rightarrow 0} (\textbf{s}_1;\textbf{s}_1')\\
&=\gamma ^\mathrm{T}_{0\rightarrow m} (\textbf{s}_2;\textbf{s}_2')\left[\gamma ^\mathrm{T}_{0\rightarrow m} (\textbf{s}_1';\textbf{s}_1)\right]^* 
\end{align*}
i.e.,
\begin{align*}
R_m^<(\textbf{s}_1,\textbf{s}_2;\textbf{s}_1',\textbf{s}_2') 
&=\braket{\psi _0|\hat{\Psi}^\dag (\textbf{s}_2')\hat{\Psi}(\textbf{s}_2)|\psi _m}\\&\times\braket{\psi _0|\hat{\Psi}^\dag (\textbf{s}_1)\hat{\Psi}(\textbf{s}_1')|\psi _m}^*.
\end{align*}
In our finite-dimensional one-particle-state model, $R_m^<$ in the natural-transition orbitals basis reads
\begin{align*}
R_m^<(\textbf{s}_1,\textbf{s}_2;\textbf{s}_1',\textbf{s}_2') 
&= \sum _{p=1}^K \lambda^{0\rightarrow m}_p \,l^{0\rightarrow m}_p(\textbf{s}_2)\left[r_p^{0\rightarrow m}(\textbf{s}_2')\right]^* \\
& \times \sum _{v=1}^K \lambda^{m\rightarrow 0}_v \,l^{m\rightarrow 0}_v(\textbf{s}_1)\left[r_v^{m\rightarrow 0}(\textbf{s}_1')\right]^*.
\end{align*}
Similarly, for $R_m^>$ we identify the residue of the $m^\mathrm{th}$ term of $\chi ^>$ corresponding to the $(\Omega _{m\rightarrow 0}+ \mathrm{i}\eta ^+)$ pole as
\begin{align*}
R_m^>(\textbf{s}_1,\textbf{s}_2;\textbf{s}_1',\textbf{s}_2') &= \gamma ^\mathrm{T}_{0\rightarrow m} (\textbf{s}_1;\textbf{s}_1')\gamma ^\mathrm{T}_{m\rightarrow 0} (\textbf{s}_2;\textbf{s}_2')\\
&=\left[\gamma ^\mathrm{T}_{m\rightarrow 0} (\textbf{s}_1';\textbf{s}_1)\right]^*\gamma ^\mathrm{T}_{m\rightarrow 0} (\textbf{s}_2;\textbf{s}_2'),
\end{align*}
i.e.,
\begin{align*}
R_m^>(\textbf{s}_1,\textbf{s}_2;\textbf{s}_1',\textbf{s}_2') 
&=\braket{\psi _m|\hat{\Psi}^\dag (\textbf{s}_1)\hat{\Psi}(\textbf{s}_1')|\psi _0}^*\\ &\times \braket{\psi _m|\hat{\Psi}^\dag (\textbf{s}_2')\hat{\Psi}(\textbf{s}_2)|\psi _0}.
\end{align*}
In our finite-dimensional one-particle-state model, $R_m^>$ in the natural-transition orbitals basis reads
\begin{align*}
R_m^>(\textbf{s}_1,\textbf{s}_2;\textbf{s}_1',\textbf{s}_2') 
&= \sum _{p=1}^K \lambda^{0\rightarrow m}_p \,l^{0\rightarrow m}_p(\textbf{s}_1)\left[r_p^{0\rightarrow m}(\textbf{s}_1')\right]^* \\
& \times \sum _{v=1}^K \lambda^{m\rightarrow 0}_v \,l^{m\rightarrow 0}_v(\textbf{s}_2)\left[r_v^{m\rightarrow 0}(\textbf{s}_2')\right]^*.
\end{align*}
We see that, in $R_m^<(\textbf{s}_1,\textbf{s}_2;\textbf{s}_1',\textbf{s}_2') $, the  $\textbf{s}_1'$ and $\textbf{s}_2$ coordinates are associated with the probing of one body in $\ket{\psi _m}$. Tracing out this contribution gives
\begin{align*}
\gamma ^r_{0\rightarrow m}(\textbf{s}_1;\textbf{s}_2') = \int _{S_4} \!\!\!\mathrm{d}\textbf{s}_1' \int _{S_4} \!\!\!\mathrm{d}\textbf{s}_2 \, \delta (\textbf{s}_1'-\textbf{s}_2) R_m^<(\textbf{s}_1,\textbf{s}_2;\textbf{s}_1',\textbf{s}_2'),
\end{align*} 
whose expression in our finite-dimensional one-particle-state basis model is
\begin{align*}
\gamma ^r_{0\rightarrow m}(\textbf{s}_1;\textbf{s}_2') &= \sum _{p=1}^K \sum _{q=1}^K  \left(\bm{\gamma}^r_{0\rightarrow m} \right)_{p,q}k_p(\textbf{s}_1)k_q^*(\textbf{s}_2')\\
&= \sum _{p=1}^K \left(\lambda ^{0 \rightarrow m}_p\right)^2 r_p^{0\rightarrow m}(\textbf{s}_1)\left[r_p^{0\rightarrow m}(\textbf{s}_2')\right]^*.  
\end{align*}
Conversely, the $\textbf{s}_1'$ and $\textbf{s}_2$ coordinates are associated in $R^>_m$ with the probing of one body in $\ket{\psi _0}$. Tracing out this contribution gives
\begin{align*}
\gamma ^l_{0\rightarrow m}(\textbf{s}_1;\textbf{s}_2') = \int _{S_4} \!\!\!\mathrm{d}\textbf{s}_1' \int _{S_4} \!\!\!\mathrm{d}\textbf{s}_2 \, \delta (\textbf{s}_1'-\textbf{s}_2) R_m^>(\textbf{s}_1,\textbf{s}_2;\textbf{s}_1',\textbf{s}_2'),
\end{align*} 
whose expression in our finite-dimensional one-particle-state basis model is
\begin{align*}
\gamma ^l_{0\rightarrow m}(\textbf{s}_1;\textbf{s}_2') &= \sum _{p=1}^K \sum _{q=1}^K  \left(\bm{\gamma}^l_{0\rightarrow m} \right)_{p,q}k_p(\textbf{s}_1)k_q^*(\textbf{s}_2') \\
&= \sum _{p=1}^K \left(\lambda ^{0 \rightarrow m}_p\right)^2 l^{0 \rightarrow m}_p(\textbf{s}_1)\left[l_p^{0 \rightarrow m}(\textbf{s}_2')\right]^*. 
\end{align*}
\subsubsection{Densities}
\noindent We introduce the one-body reduced detachment, attachment, $l-$ and $r-$density functions $n_{\ell \rightarrow m}^{\omega,\textbf{s}}$ (with $\omega \in \left\lbrace d, a,l,r\right\rbrace$ and $(\ell , m) \in S^2$), which map $S_4$ to $\mathbb{R}_+$ as
\begin{align*}
 \textbf{s}_1 \longmapsto n_{\ell \rightarrow m}^{\omega,\textbf{s}} (\textbf{s}_1) = \sum _{r=1}^K \sum _{s=1}^K \left(\bm{\gamma}^\omega_{\ell \rightarrow m}\right)_{r,s} k _r(\textbf{s}_1)k _s^*(\textbf{s}_1). 
\end{align*}
The fact that these four maps define nonnegative-valued functions is justified by the fact that any of these maps can be written in a basis which makes its coefficient matrix diagonal and positive semidefinite, so the corresponding function is a sum of products between nonnegative coefficients and the squared modulus of a spinorbital value. For visualization purposes, the spin variable can be summed out, as described in section \ref{sec:AppRDMs}.
\section{Interpretation}\label{sec:interpretation}
\noindent Providing a physical interpretation to the tools related above is a very complicated task — sometimes impossible. Indeed, there is, with such objects, the constant risk to overinterpret their nature, and we sometimes try to capture physical insights from maps that are constructed for practical purposes (reducing the number of terms of an expectation value using natural orbitals), and which are used to derive physical information \textit{a posteriori} (see below). The \textit{state} one-electron reduced density kernel is interpreted as the time-independent two-point field correlation function corresponding to a quantum electronic state — what happens at a certain point in $S_4$ is correlated with what happens at another point. It is useful for deriving the expression for the expectation value of non-local operators. On the other hand, the one-body reduced transition density kernel is a much more complex object. It seems difficult — if possible — to provide a universal statistical/physical interpretation to it if we consider two general many-body wavefunctions. It is identified in many contributions as an exciton (bound hole-electron pair) wavefunction. Consistently with the exciton-wavefunction interpretation of the one-body reduced transition density kernel, $\gamma^l$ and $\gamma^r$ are often identified as transition-electron and transition-hole densities respectively. They are seen as resulting from tracing out the hole/electron contribution from $R^>$ and $R^<$ which are both understood as the product of the exciton wavefunction value by the complex conjugate of another of its values. The procedure suggested is the same as the one we use for deriving the state one-body reduced density kernel using partial trace — see \eqref{eq:partial_trace_integral} and \eqref{eq:TotalKernel} with $\lambda = \mathrm{T}$ and $\ell = m$. The use of this model is discussed below in more details. While the transition density kernel already seems difficult to understand, $R^<$ and $R^>$ are even more complicated objects — what is the physical content of a product of two two-point, two-state cross-correlation functions? While the interpretation of $\hat{\mathrm{T}}_N^{\ell \rightarrow m}(\hat{\mathrm{T}}_N^{\ell \rightarrow m})^\dag$ and $(\hat{\mathrm{T}}_N^{\ell \rightarrow m})^\dag\hat{\mathrm{T}}_N^{\ell \rightarrow m}$ is highly straightforward — these two operator products are $N$-electron state-projectors, the former projects onto $\ket{\psi_m}$, the latter projects onto $\ket{\psi _\ell}$ — the $l-$ and $r-$density matrices for their part are much more complex: they are the matrix representation of some
$\hat{\gamma}^\mathrm{T}_{\ell\rightarrow m}\!\! \left(\hat{\gamma}^\mathrm{T}_{\ell\rightarrow m}\right)\!^\dag$
and
$\left(\hat{\gamma}^\mathrm{T}_{\ell\rightarrow m}\right)^{\!\dag}\!\hat{\gamma}^\mathrm{T}_{\ell\rightarrow m} $
one-electron operators mapping $\mathcal{K}$ onto itself. What are these operators? When acting on a one-electron quantum state, what is the nature of the one-electron state that is obtained, and how can we use it? 

The eigenvectors of the $l-$ and $r-$density matrices can be used to rewrite the residues of $R^<$ and $R^>$ using $K^2$ terms rather than $K^4$, and to compute transition moments (and, more generally, one-body transition properties, especially those corresponding to differential operators) with $K$ terms instead of $K^2$. 

Spinorbitals are often seen as one-electron boxes, and used as such for qualitative interpretation purposes. Since left- and right-singular vectors of the one-body reduced transition density matrix are the eigenvectors of the $l$-density matrix (sometimes termed transition-electron density matrix) and $r$-density matrix (sometimes termed transition-hole density matrix), the left and right natural transition orbitals are often termed ``transition-electron'' (or more simply ``electron'') and ``transition-hole'' (or more simply ``hole'') natural transition orbitals, respectively. They are also given some physical interpretation, and the electronic transition picture is often approximated using a collection of ``transition-hole natural transition orbital $\longrightarrow$ transition-electron natural transition orbital'' representations, together with the singular values corresponding to the displayed natural transition orbital pairs. 

The vocables ``occupied/virtual natural transition orbitals'' — occupied (respectively, virtual) for the hole (respectively, electron) natural transition orbitals — are also commonly used in the literature \cite{etienne_fluorene-imidazole_2016}. These vocables are inherited from the historical introduction of these tools with the departure state being the Fermi vacuum, with two separate spaces of spinorbitals, and the arrival state being a linear combination of Fock states with single-electron/single-orbital replacement relatively to Fermi vacuum — the so-called configuration interaction singles method. However, since any hole and any electron natural transition orbital is a linear combination of all the orbitals from the one-electron-state basis, it naturally comes that when the electronic transition takes place between two multi-determinantal states, these vocables are not meaningful anymore. It is essentially misleading anyways since it suggests that the occupied (respectively, virtual) natural transition orbitals are occupied (respectively, unoccupied) in the departure (respectively, arrival) state, and it strongly suggests a picture of the electronic transition with departure and arrival orbitals. This model for the transition and use of the natural transition orbitals is solely meaningful and complete in the configuration interaction singles approximation, in which the natural difference and transition orbitals have identical expressions \cite{etienne_auxiliary_2021}.

Consistently with what precedes, the transition-hole and transition-electron densities are often used for providing a ``transition-hole density $\longrightarrow$ transition-electron density'' picture of the electronic transition in a simplified one-hole-one-electron reduced model for the electronic transition. Such a picture strongly suggests a departure and an arrival for the densities, while we will show in the following lines that the two corresponding densities are more related to a density-density \textit{coupling}. This term is preferred since the transition-hole and the transition-electron densities do not necessarily integrate to unity, though being derived from the residues of a single-hole-single-electron coupling function. In a time-dependent picture of the light-induced transition process, the $N$-body transition operator is met when the two stationary states which compose it — the eigenstates corresponding to the reference (or zeroth-order) Hamiltonian — are not eigenstates of the field-perturbed Hamiltonian, and is associated to \textit{coherences}, i.e., interferences effects — how the two states can be coupled upon the interaction of an electromagnetic field with the molecular system. Tracing out ($N-1$)-electron contributions leaves the one-electron transition density matrix containing the one-body information related to this coupling. A proper use of the natural transition orbitals should involve no directionality (a double arrow ``$\longleftrightarrow$'' might be more meaningful than a single-directional arrow ``$\longrightarrow$''), for the hole and particle entities are described as bound, and evolving together rather than being part of a departure-arrival situation. The associated densities can also be regarded as the two sides of a same coin, and are associated to the first-order (i.e., linear) density response — the propagation of a two-component density perturbation. 
 
Note also that though the first-order response framework is sufficient for recovering the exact transition energies \cite{strinati_effects_1984}, it does not mean that the information captured into the residues of the bound one-hole-one-electron pair correlation function regarding the electronic-structure reorganization is complete. Consider for instance the transition between two multideterminantal states. Unlike the detachment/attachment density picture of the transition, which contains all the information that can be condensed into one-body density functions from many-body states — excitation degrees up to $N$ can in principle be included into the difference density matrix —, the one-electron reduced transition density kernel and matrix solely retain contributions coupling Fock-state manifolds differing by at most one value among the spinorbital occupation numbers: if the departure state contains at most the $p-$excitation manifold and the arrival state contains at most the $z-$excitation manifold ($p$ and $z$ are two nonnegative integers lower or equal to $N$; these excitation degrees are given relatively to a Fermi vacuum reference common to the departure and arrival states), every couple of manifolds characterized by a value superior to unity for the absolute value of the difference in excitation degree will not be accounted for in the one-electron transition density picture, including natural transition orbital, $l$- and $r-$density representations. In summary, if $(| p - z | > 1)$, information will be necessarily lost. Orbital relaxation effects are also absent from the one-electron transition density matrix, natural transition orbital, and $l-$ and $r-$density analyses, while they are included in the detachment/attachment picture.
 
What is more, the picture of the one-body reduced transition density matrix containing the single-electron replacement coefficients (i.e., the single-excitation coefficients in a full-configuration interaction picture) to the electronic transition so that its configuration interaction singles component can be pictured as a linear combination of hole-to-electron natural transition orbital population transfer is also not universal. A counterexample is given in ref. \cite{etienne_auxiliary_2021} where, using a \textit{reductio ad absurdum} argument, we show that the transition density matrix issued from the construction of corresponding auxiliary many-body wavefunction differs from the one used for building the auxiliary many-body wavefunction itself.

On the other hand, state one-electron densities can be physically measured, and we have the property that difference between the attachment and detachment density functions is equal to the one-electron difference density whose negative/positive values — unlike those of the one-electron reduced transition density — depict actual density transfer, i.e., neat density depletion/accumulation zones, so that the detachment/attachment tool provides a ``simplified'' picture of the density rearrangement — the contribution of the two states to the electronic-structure difference. Yet, though the detachment and attachment densities are one-body density functions, they do not necessarily integrate to unity, so they are not necessarily picturing a single-electron evolution (\textit{vide supra}). The detachment/attachment picture is strongly directional, and suggests a sort of density ``departure'' and ``arrival''. It is meant to \textit{compare} the electronic structure of two eigenstates of the reference Hamiltonian — the one in which the system is \textit{before} light-matter interaction, and the one in which the system is \textit{after} the light-matter interaction, hence the terms departure/arrival states. Charge \textit{transfer} can be visualized when \textit{comparing} the states using the difference density, or the detachment and attachment densities as ``detachment density $\longrightarrow$ attachment density'' which allows the identification of the nature of the transition itself — what are the consequences of state transfer for the electronic structure of the system.

\section{Conclusion}
\noindent Every left natural transition orbital is always paired to a right natural transition orbital, but these pairs are not meant to be interpreted as departure/arrival orbitals in the general case. On the other hand, the natural difference orbitals are not universally paired. In other words, there is no universal departure/arrival natural-orbital representation of electronic transitions that can be derived from the one-electron reduced difference or transition density matrix. 
\newpage
\section*{Acknowledgements}
\noindent Benjamin Lasorne, Jean Christophe Tremblay, Sébastien Lebègue, Felix Plasser, Jérôme Dorignac, Yohann Scribano, Christophe Raynaud and Matthieu Saubanère are gratefully acknowledged for interesting and insightful discussions on the topic. Jérémy Morere, Guillaume Wagner and Charles Willig are also acknowledged for proofreading the manuscript. 

\bibliographystyle{ieeetr}

\begin{thebibliography}{10}

\bibitem{maurice_configuration_1995-1}
D.~Maurice and M.~Head-Gordon, ``Configuration interaction with single
  substitutions for excited states of open-shell molecules,'' {\em
  International Journal of Quantum Chemistry}, vol.~56, pp.~361--370, Feb.
  1995.

\bibitem{david_sherrill_configuration_1999}
C.~David~Sherrill and H.~F. Schaefer, ``The {Configuration} {Interaction}
  {Method}: {Advances} in {Highly} {Correlated} {Approaches},'' in {\em
  Advances in {Quantum} {Chemistry}}, vol.~34, pp.~143--269, Elsevier, 1999.

\bibitem{sekino_linear_1984}
H.~Sekino and R.~J. Bartlett, ``A linear response, coupled-cluster theory for
  excitation energy,'' {\em International Journal of Quantum Chemistry},
  vol.~26, pp.~255--265, Mar. 1984.

\bibitem{koch_coupled_1990}
H.~Koch and P.~Jo/rgensen, ``Coupled cluster response functions,'' {\em The
  Journal of Chemical Physics}, vol.~93, pp.~3333--3344, Sept. 1990.
\newblock Publisher: American Institute of Physics.

\bibitem{hirata_configuration_1999}
S.~Hirata, M.~Head-Gordon, and R.~J. Bartlett, ``Configuration interaction
  singles, time-dependent {Hartree}–{Fock}, and time-dependent density
  functional theory for the electronic excited states of extended systems,''
  {\em The Journal of Chemical Physics}, vol.~111, pp.~10774--10786, Dec. 1999.
\newblock Publisher: American Institute of Physics.

\bibitem{casida_time-dependent_1995}
M.~E. Casida, ``Time-{Dependent} {Density} {Functional} {Response} {Theory} for
  {Molecules},'' in {\em Recent {Advances} in {Density} {Functional}
  {Methods}}, vol.~Volume 1 of {\em Recent {Advances} in {Computational}
  {Chemistry}}, pp.~155--192, WORLD SCIENTIFIC, Nov. 1995.

\bibitem{ziegler_derivation_2014}
T.~Ziegler, M.~Krykunov, and J.~Autschbach, ``Derivation of the {RPA} ({Random}
  {Phase} {Approximation}) {Equation} of {ATDDFT} ({Adiabatic} {Time}
  {Dependent} {Density} {Functional} {Ground} {State} {Response} {Theory}) from
  an {Excited} {State} {Variational} {Approach} {Based} on the {Ground} {State}
  {Functional},'' {\em Journal of Chemical Theory and Computation}, vol.~10,
  pp.~3980--3986, Sept. 2014.

\bibitem{fromager_individual_2020}
E.~Fromager, ``Individual correlations in ensemble density-functional theory:
  {State}-driven/density-driven decompositions without additional {Kohn}-{Sham}
  systems,'' {\em Physical Review Letters}, vol.~124, p.~243001, June 2020.
\newblock arXiv: 2001.08605.

\bibitem{pernal_time-dependent_2007-1}
K.~Pernal, O.~Gritsenko, and E.~J. Baerends, ``Time-dependent
  density-matrix-functional theory,'' {\em Physical Review A}, vol.~75,
  p.~012506, Jan. 2007.

\bibitem{rebolini_electronic_2013}
E.~Rebolini, J.~Toulouse, and A.~Savin, ``Electronic excitation energies of
  molecular systems from the {Bethe}-{Salpeter} equation: {Example} of the
  {H$_2$} molecule,'' {\em arXiv:1304.1314 [cond-mat, physics:physics]}, Apr.
  2013.
\newblock arXiv: 1304.1314 version: 1.

\bibitem{leng_gw_2018}
X.~Leng, F.~Jin, M.~Wei, and Y.~Ma, ``{GW} method and {Bethe}–{Salpeter}
  equation for calculating electronic excitations,'' {\em Wiley
  Interdisciplinary Reviews: Computational Molecular Science}, vol.~6,
  pp.~532--550, July 2018.

\bibitem{oddershede_polarization_1978}
J.~Oddershede, ``Polarization {Propagator} {Calculations},'' in {\em Advances
  in {Quantum} {Chemistry}} (P.-O. Löwdin, ed.), vol.~11, pp.~275--352,
  Academic Press, Jan. 1978.

\bibitem{strinati_application_1988}
G.~Strinati, ``Application of the {Green}’s functions method to the study of
  the optical properties of semiconductors,'' {\em La Rivista del Nuovo Cimento
  (1978-1999)}, vol.~11, pp.~1--86, Dec. 1988.

\bibitem{bappler_exciton_2014}
S.~A. Bäppler, F.~Plasser, M.~Wormit, and A.~Dreuw, ``Exciton analysis of
  many-body wave functions: {Bridging} the gap between the quasiparticle and
  molecular orbital pictures,'' {\em Physical Review A}, vol.~90, p.~052521,
  Nov. 2014.

\bibitem{dreuw_single-reference_2005}
A.~Dreuw and M.~Head-Gordon, ``Single-{Reference} ab {Initio} {Methods} for the
  {Calculation} of {Excited} {States} of {Large} {Molecules},'' {\em Chemical
  Reviews}, vol.~105, pp.~4009--4037, Nov. 2005.
\newblock Publisher: American Chemical Society.

\bibitem{luzanov_charge_1978}
A.~V. Luzanov, ``Charge transfer and localization during electronic excitation
  of molecules,'' {\em Theoretical and Experimental Chemistry}, vol.~13,
  pp.~433--440, Sept. 1978.

\bibitem{plasser_detailed_2017}
F.~Plasser, S.~A. Mewes, A.~Dreuw, and L.~González, ``Detailed {Wave}
  {Function} {Analysis} for {Multireference} {Methods}: {Implementation} in the
  {Molcas} {Program} {Package} and {Applications} to {Tetracene},'' {\em
  Journal of Chemical Theory and Computation}, vol.~13, pp.~5343--5353, Nov.
  2017.
\newblock Publisher: American Chemical Society.

\bibitem{luzanov_structure_1980}
A.~V. Luzanov, ``The {Structure} of the {Electronic} {Excitation} of
  {Molecules} in {Quantum}-chemical {Models},'' {\em Russian Chemical Reviews},
  vol.~49, p.~1033, Nov. 1980.
\newblock Publisher: IOP Publishing.

\bibitem{luzanov_analysis_2006}
A.~V. Luzanov and O.~V. Prezhdo, ``Analysis of multiconfigurational wave
  functions in terms of hole-particle distributions,'' {\em The Journal of
  Chemical Physics}, vol.~124, p.~224109, June 2006.

\bibitem{ronca_charge-displacement_2014}
E.~Ronca, M.~Pastore, L.~Belpassi, F.~De~Angelis, C.~Angeli, R.~Cimiraglia, and
  F.~Tarantelli, ``Charge-displacement analysis for excited states,'' {\em The
  Journal of Chemical Physics}, vol.~140, p.~054110, Feb. 2014.

\bibitem{li_particlehole_2015}
Y.~Li and C.~A. Ullrich, ``The {Particle}–{Hole} {Map}: {A} {Computational}
  {Tool} {To} {Visualize} {Electronic} {Excitations},'' {\em Journal of
  Chemical Theory and Computation}, vol.~11, pp.~5838--5852, Dec. 2015.

\bibitem{etienne_charge_2019-1}
T.~Etienne and M.~Pastore, ``Charge separation: {From} the topology of
  molecular electronic transitions to the dye/semiconductor interfacial
  energetics and kinetics,'' {\em arXiv}, 2019.
\newblock arXiv: 1811.10526.

\bibitem{monino_upper_2021}
E.~Monino and T.~Etienne, ``Upper bound for the charge transferred during a
  molecular electronic transition: insights from matrix analysis,'' {\em
  arXiv:2104.13465 [physics]}, Apr. 2021.
\newblock arXiv: 2104.13465.

\bibitem{etienne_toward_2014}
T.~Etienne, X.~Assfeld, and A.~Monari, ``Toward a {Quantitative} {Assessment}
  of {Electronic} {Transitions}’ {Charge}-{Transfer} {Character},'' {\em
  Journal of Chemical Theory and Computation}, vol.~10, pp.~3896--3905, Sept.
  2014.

\bibitem{plasser_new_2014-2}
F.~Plasser, M.~Wormit, and A.~Dreuw, ``New tools for the systematic analysis
  and visualization of electronic excitations. {I}. {Formalism},'' {\em The
  Journal of Chemical Physics}, vol.~141, p.~024106, July 2014.
\newblock Publisher: American Institute of Physics.

\bibitem{breuil_diagnosis_2019-2}
G.~Breuil, K.~Shehu, E.~Lognon, S.~Pitié, B.~Lasorne, and T.~Etienne,
  ``Diagnosis of two evaluation paths to density-based descriptors of molecular
  electronic transitions,'' {\em arXiv}, 2019.
\newblock arXiv: 1902.05840.

\bibitem{luzanov_interpretation_1980}
A.~V. Luzanov and V.~F. Pedash, ``Interpretation of excited states using
  charge-transfer numbers,'' {\em Theoretical and Experimental Chemistry},
  vol.~15, pp.~338--341, July 1980.

\bibitem{ciofini_through-space_2012}
I.~Ciofini, T.~Le~Bahers, C.~Adamo, F.~Odobel, and D.~Jacquemin,
  ``Through-{Space} {Charge} {Transfer} in {Rod}-{Like} {Molecules}: {Lessons}
  from {Theory},'' {\em The Journal of Physical Chemistry C}, vol.~116,
  pp.~11946--11955, June 2012.

\bibitem{head-gordon_analysis_1995}
M.~Head-Gordon, A.~M. Grana, D.~Maurice, and C.~A. White, ``Analysis of
  {Electronic} {Transitions} as the {Difference} of {Electron} {Attachment} and
  {Detachment} {Densities},'' {\em The Journal of Physical Chemistry}, vol.~99,
  pp.~14261--14270, Sept. 1995.

\bibitem{etienne_comprehensive_2020-1}
T.~Etienne, ``A comprehensive, self-contained derivation of the one-body
  density matrices from single-reference excited-state calculation methods
  using the equation-of-motion formalism,'' {\em arXiv}, 2020.
\newblock arXiv: 1811.08849.

\bibitem{luzanov_application_1976}
A.~V. Luzanov, A.~A. Sukhorukov, and V.~E. Umanskii, ``Application of
  transition density matrix for analysis of excited states,'' {\em Theoretical
  and Experimental Chemistry}, vol.~10, pp.~354--361, July 1976.

\bibitem{martin_natural_2003}
R.~L. Martin, ``Natural transition orbitals,'' {\em The Journal of Chemical
  Physics}, vol.~118, pp.~4775--4777, Mar. 2003.

\bibitem{strinati_effects_1984}
G.~Strinati, ``Effects of dynamical screening on resonances at inner-shell
  thresholds in semiconductors,'' {\em Physical Review B}, vol.~29,
  pp.~5718--5726, May 1984.
\newblock Publisher: American Physical Society.

\bibitem{etienne_fluorene-imidazole_2016}
T.~Etienne, H.~Gattuso, C.~Michaux, A.~Monari, X.~Assfeld, and E.~A. Perpète,
  ``Fluorene-imidazole dyes excited states from first-principles
  calculations—{Topological} insights,'' {\em Theoretical Chemistry
  Accounts}, vol.~135, p.~111, Apr. 2016.

\bibitem{etienne_auxiliary_2021}
T.~Etienne, ``Auxiliary many-body wavefunctions for {TDDFRT} electronic excited
  states: {Consequences} for the representation of molecular electronic
  transitions,'' {\em arXiv:2104.13616 [physics]}, Apr. 2021.
\newblock arXiv: 2104.13616.

\end{thebibliography}

\end{document}